\documentclass[12pt]{article}
\usepackage{amsmath,amssymb,bm,bbm,mathrsfs,amscd}
\usepackage{calc}
\usepackage{color}
\usepackage[margin=2.5cm]{geometry}
\usepackage{amsthm,placeins,pdfpages}
\usepackage{xurl}
\usepackage{hyperref,cite}
\usepackage{graphicx}

 \newcommand{\house}{\mathrm{H}}
\newcommand{\work}{\mathrm{W}}
\newcommand{\school}{\mathrm{S}}
\newcommand{\retire}{\mathrm{Rh}}
\newcommand{\hsp}{\mathrm{Hsp}}

\definecolor{codegreen}{rgb}{0,0.6,0}
\definecolor{codegray}{rgb}{0.5,0.5,0.5}
\definecolor{codepurple}{rgb}{0.58,0,0.82}
\definecolor{backcolour}{rgb}{0.95,0.95,0.92}

\def\la{12/21/2020} 

\begin{document}

\date{}

\title{High-resolution agent-based modeling of COVID-19 spreading in a small town}

\maketitle

\author{\large Agnieszka Truszkowska$^1$, Brandon Behring$^1$, Jalil Hasanyan$^1$, Lorenzo Zino$^2$, Sachit Butail$^3$, Emanuele Caroppo$^{4,5}$, Zhong-Ping Jiang$^6$, Alessandro Rizzo$^{7,8}$, and Maurizio Porfiri$^{1,9,10}$}

\date{\vspace{.5cm}

\normalsize 
$^1$Department of Mechanical and Aerospace Engineering and Department of Biomedical Engineering, New York University Tandon School of Engineering, Brooklyn NY 11201, USA\\
$^2$Faculty of Science and Engineering, University of Groningen, 9747 AG Groningen, Netherlands\\
$^3$Department of Mechanical Engineering, Northern Illinois University, DeKalb IL 60115, USA\\
$^4$Mental Health Department, Local Health Unit ROMA 2, 00174 Rome, Italy\\
$^5$University Research Center He.R.A., Università Cattolica del Sacro Cuore, 00168 Rome, Italy\\
$^6$Department of Electrical and Computer Engineering, New York University Tandon School of Engineering, 370 Jay Street, Brooklyn NY 11201, USA\\
$^7$Department of Electronics and Telecommunications, Politecnico di Torino, 10129 Turin, Italy\\
$^8$Office of Innovation, New York University Tandon School of Engineering, Brooklyn NY 11201, USA\\
$^9$Department of Biomedical Engineering, New York University Tandon School of Engineering, Brooklyn NY 11201, USA\\
$^{10}$Center for Urban Science and Progress, Tandon School of Engineering, New York University, 370 Jay Street, Brooklyn, NY 11201, USA\\\vspace{.2cm}

\noindent Correspondence should be addressed to: \url{mporfiri@nyu.edu}}


\newpage



\begin{abstract}
Amid the ongoing COVID-19 pandemic, public health authorities and the general population are striving to achieve a balance between safety and normalcy. Ever changing conditions call for the development of theory and simulation tools to finely describe multiple strata of society while supporting the evaluation of ``what-if’' scenarios. Particularly important is to assess the effectiveness of potential testing approaches and vaccination strategies. Here, an agent-based modeling platform is proposed to simulate the spreading of COVID-19 in small towns and cities, with a single-individual resolution. The platform is validated on real data from New Rochelle, NY ---one of the first outbreaks registered in the United States. Supported by expert knowledge and informed by reported data, the model incorporates detailed elements of the spreading within a statistically realistic population. Along with pertinent functionality such as testing, treatment, and vaccination options, the model accounts for the burden of other illnesses with symptoms similar to COVID-19. Unique to the model is the possibility to explore different testing approaches ---in hospitals or drive-through facilities--- and vaccination strategies that could prioritize vulnerable groups. Decision making by public authorities could benefit from the model, for its fine-grain resolution, open-source nature, and wide range of features.\end{abstract}

\section{Introduction}
In December 2019, COVID-19 was first observed in humans in Wuhan, the Hubei Province's capital in China. The World Health Organization (WHO) declared this outbreak as a Public Health Emergency of International Concern on January 30, 2020, and later named it a pandemic on March 11\textsuperscript{th}, 2020. As of December 23\textsuperscript{rd}, 2020, the WHO has reported 76,382,044 cases globally, with 1,702,128 deaths.\textsuperscript{\cite{WHO}} In the US, the number of infected individuals keeps rising, with tens of thousands of newly infected cases discovered every day. The Centers for Disease Control and Prevention (CDC) has reported 17,974,303 cases as of December 23\textsuperscript{rd}, 2020.\textsuperscript{\cite{CDC}} Following an unprecedented containment campaign of lockdowns, most countries seek a delicate balance between safety and normalcy, aiming for a safe return to normal activities amidst less restrictive conditions.

Timely case detection through efficient testing and contact tracing is among the key components required for lowering the COVID-19 spread before a vaccine becomes available.\textsuperscript{\cite{unwin2020report, aleta2020modeling, salathe2020covid,reynalara2020virus}} Important questions on testing pertain to the identification of infected individuals and their contacts. Addressing these questions calls for an improved understanding of community structure, outbreak locations, and individual lifestyles.\textsuperscript{\cite{heidarzadeh2020two,singh2020partially}} Due to the  scale of the COVID-19 epidemic, an additional burden has been placed on traditional testing sites, such as hospitals, emergency rooms, and walk-in clinics, thereby challenging their safety.\textsuperscript{\cite{ton2020covid,evans2020covid,lindholm2020outcomes}}  

Computational models are powerful tools for understanding novel epidemics and evaluating the effectiveness of potential countermeasures.\textsuperscript{\cite{estrada2020covid, vespignani2020modelling,baldea2020suppression,gilbert2020africa,DellaRossa2020}} Agent-based models (ABMs) are a class of computational models that provide a high-resolution---both temporal and spatial---representation of the epidemic at the individual level.\textsuperscript{\cite{aleta2020modeling,perez2009agent,Perra2015,ferguson2005strategies,ferguson2006strategies,ajelli2010comparing}} These models afford consideration of multiple physical locations, such as businesses or schools, as well as unique features of communities, like human behavioral trends or local mobility patterns. 

Once validated, ABMs can be used to test competing ``what-if" scenarios that would otherwise require impractical and, potentially, unethical experiments. For example, Ferguson et al. developed an ABM to investigate the impact of non-pharmaceutical interventions for COVID-19, such as nation-wide confinement in the United Kingdom and the United States.\textsuperscript{\cite{ferguson2020impact}} Aleta et al. proposed an ABM for the entire Boston metropolitan area to elucidate the role of different types of contact tracing measures in maintaining low levels of infection.\textsuperscript{\cite{aleta2020modeling}} Gressman and Peck created an ABM of a university campus to examine strategies for the safe reopening of higher education institutions.\textsuperscript{\cite{gressman2020simulating}} Hinch et al. formulated an open-source agent-based
modeling framework to support the analysis of select non-pharmaceutical interventions and contact tracing schemes.\textsuperscript{\cite{hinch2020openabm}} The merits of ABMs have been recognized by a large number of studies, which have shed light on technical aspects of their implementation as well as their scalability across scenarios.\textsuperscript{\cite{kerr2020covasim,gopalan2020reliable,goldenbogen2020geospatial,kuzdeuov2020network,keskinocak2020impact}}

The focus of existing ABMs is either small micro-environments or entire countries and large metropolitan areas, where the population is purposefully coarsened to enable numerical simulation. Medium-size and highly resolved communities constitute an important, yet unconsidered, modeling scale for COVID-19. America is a ``nation of small towns,'' as described by the US Census,\textsuperscript{\cite{USSmallTowns}}, and after the initial wave of spreading in the New York metropolitan area, we see that small and medium-size towns are increasingly hit by the pandemic. A high-resolution ABM can closely capture real-world communities and interaction patterns at this intermediate scale, thereby allowing them to carefully reflect town-specific lifestyles without requiring coarsening to perform massive simulation campaigns. Thus, the synthetic generation of a one-to-one virtual population with its own individual buildings (residential and public) opens a wide range of new possibilities in epidemiological analysis, which may inform public health authorities to design accurate and targeted interventions. The analysis could include lockdowns of different parts of the town and afford to quantify the effect of testing practices, treatment prevalence, and vaccination strategies. 

Toward the study of these medium-size populations, we develop an agent-based modeling platform of COVID-19 for the entire town of New Rochelle, located in Westchester County in New York, US. This location was chosen as it was one of the earliest COVID-19 outbreaks in the US and is representative of a typical small town. The ABM replicates, geographically and demographically, the town structure obtained from the US Census statistics.\textsuperscript{\cite{USCensus}} The model is based on the earlier ABM developed by Ferguson's research group to study pandemic influenza,\textsuperscript{\cite{ferguson2005strategies,ferguson2006strategies,ajelli2010comparing}} which has been recently adapted to study COVID-19.\textsuperscript{\cite{ferguson2020impact}} The proposed ABM expands the original model by Ferguson et al.\textsuperscript{\cite{ferguson2005strategies,ferguson2006strategies,ajelli2010comparing}} along several directions. First, we incorporate two testing strategies: traditional, in-hospital testing with a non-negligible risk of infection, and a ``safe" drive-through testing.\textsuperscript{\cite{shah2020drive,SmartTesting, DriveTesting, EastVsWest}} Second, we account for temporal variations in time-dependent testing capacity to reflect the change in resource allocation during the course of the pandemic. Third, our model explicitly includes multiple COVID-19 treatment types, such as home isolation, hospitalization, and hospitalization in intensive care units (ICUs). Fourth, we separately track individuals with COVID-19-like symptoms due to other diseases like seasonal influenza or the common cold. These individuals are expected to play an important role in the epidemic by imposing an additional burden on testing resources. Fifth, we individually model employees in schools, hospitals, and retirement homes, enabling a dedicated consideration of these professions.  Finally, the model permits the selective study of important interventions, such as business and school closures and their reopening and vaccination strategies. 
After validation against real data, we explore alternative ``what-if'' vaccination scenarios, which may be relevant in the next several months. We consider a foreseeable situation in which a limited number of vaccines will be available. In addition to random vaccination, we explore the possibility of prioritizing vaccination for groups of individuals whose status or profession makes them particularly vulnerable to COVID-19 infection: healthcare, school, and retirement home workers, as well as residents. Along with selective vaccination of hospital employees, we assess the potential benefit of schools and businesses' concurrent closures. 

\section{The database of New Rochelle, NY}

\subsection{Geography and population data sources}
\begin{figure}
  \centering
  \includegraphics[width=0.35\textwidth]{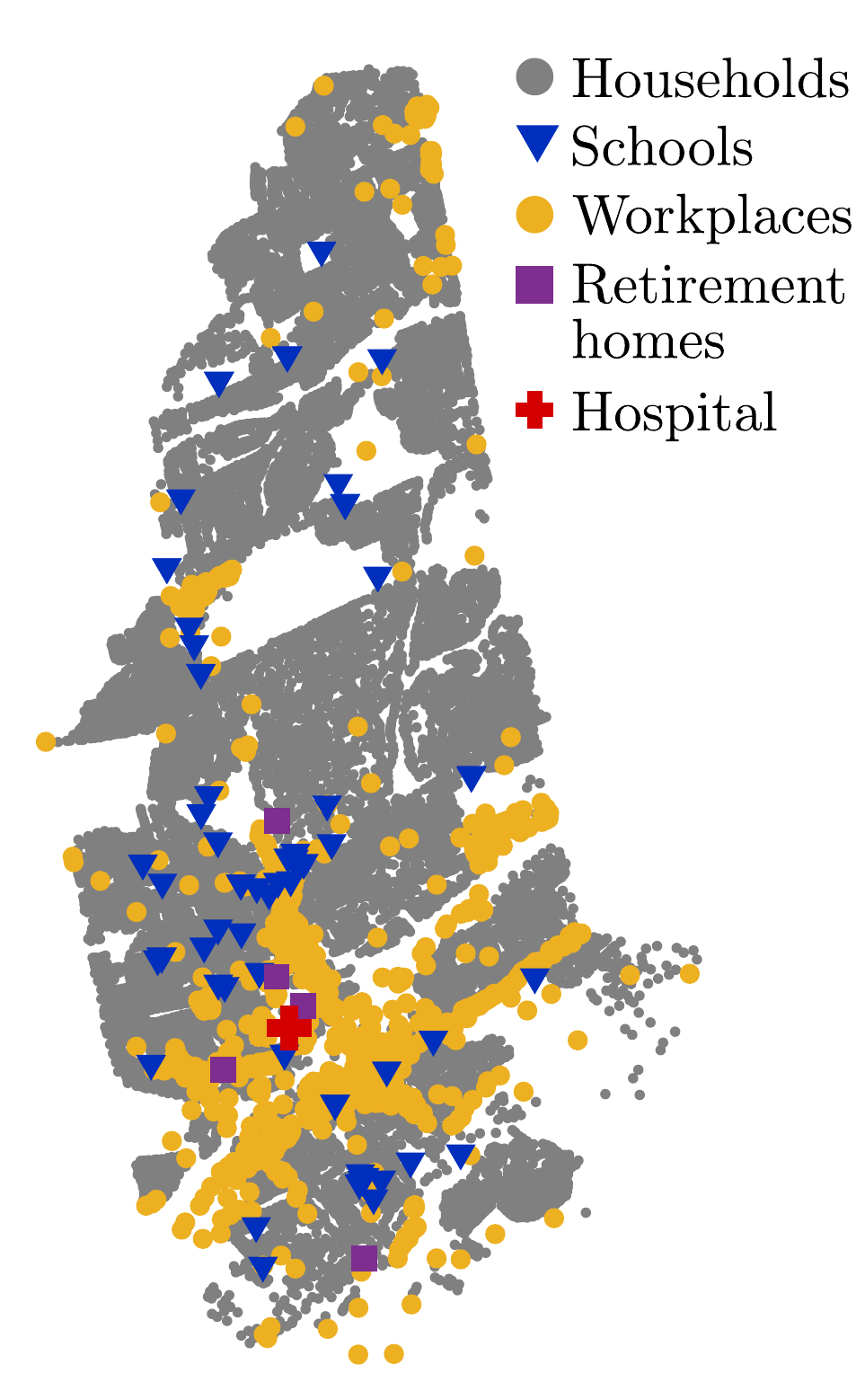}
  \caption{Map of New Rochelle, NY, which highlights the residential and public buildings  included in the database.}
  \label{fig:nrall}
\end{figure}

We collected and organized a database of geographical coordinates, type, capacity, residential buildings, and public spaces in the town of New Rochelle, NY, USA. The database was created by manually collecting geographical coordinates and characteristics of each residential and public building in the town using OpenStreetMap\textsuperscript{\cite{OpenStreetMap}} and Google Maps.\textsuperscript{\cite{GoogleMaps}} The database and the code for creating the town and its population are available through our repository (\url{https://github.com/Dynamical-Systems-Laboratory/NR-ABM-population}). The population data were collected in March and April 2020 from US Census using 2018 5-year average tables for New Rochelle.\textsuperscript{\cite{USCensus}} Since then, some of the data changed slightly, and the reader is referred to our repository for the exact datasets. Data regarding the number of students and employees in local schools was obtained from the National Center of Education Statistics,\textsuperscript{\cite{NCES}} while the hospital staff and patient number estimates were based on the records from the New York State Department of Health\textsuperscript{\cite{NYSHealth}} and the American Hospital Directory.\textsuperscript{\cite{AHD}} Figure~\ref{fig:nrall} shows the locations considered in the model.

\subsection{Household, schools, and workplace assignment of agents}

\begin{figure}
  \centering
  \includegraphics[width=0.8\textwidth]{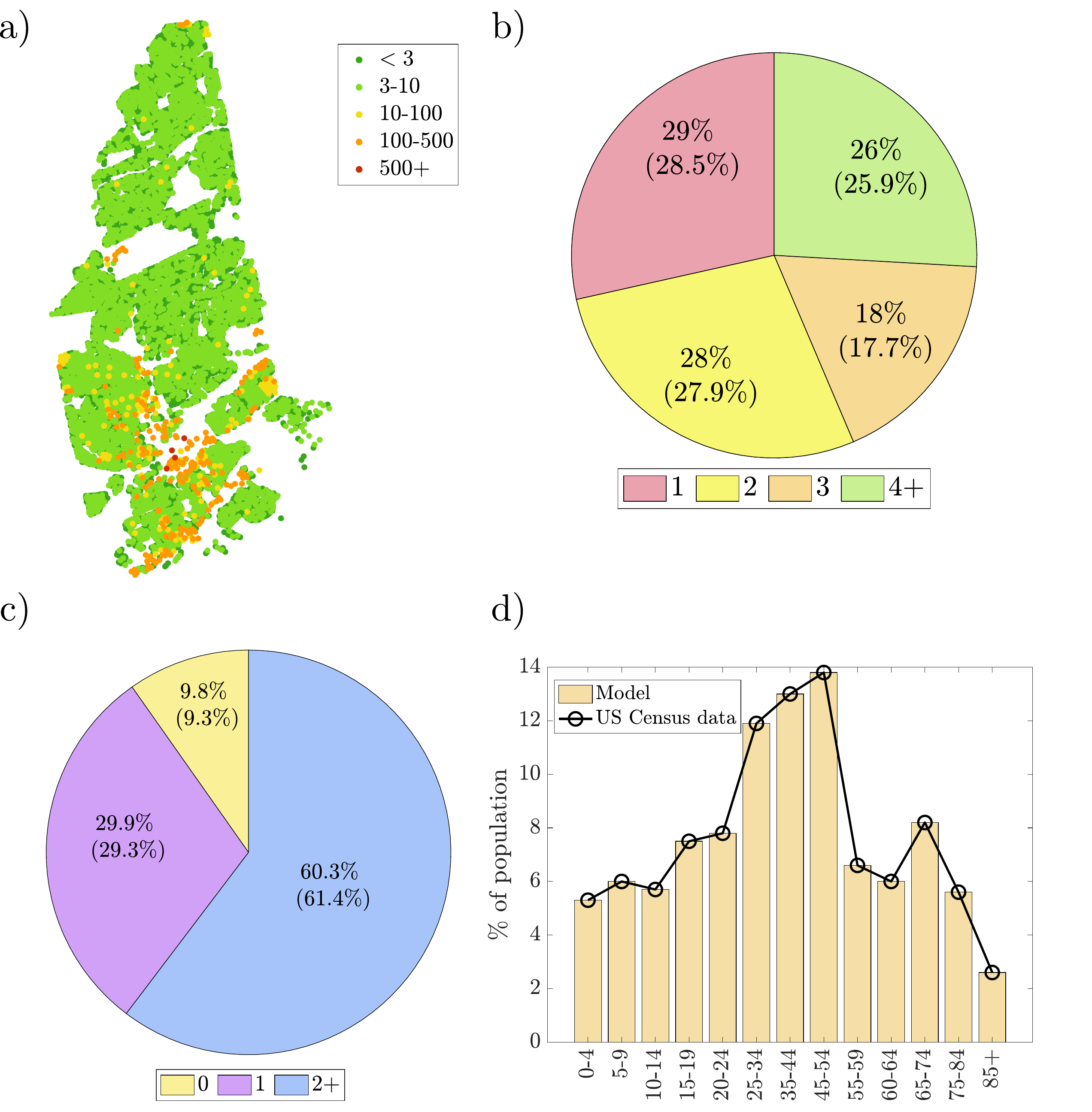}
  \caption{Select characteristics of the  created (virtual) households: a) sizes of residential buildings across town; b) percentage of households of a given size (Census data in brackets); c) distribution of employed members per family (Census data in brackets); and d) age distribution of the population.}
  \label{fig:nrstats}
\end{figure}

Households were assigned to agents using Census data on household and family structure, vacancy rate, and the number of sub-units and floors in multi-unit buildings. Figure~\ref{fig:nrstats}{\color{red}a)} shows the population distribution of the model set at the start of the simulation (see Section S3 of the Supporting Information), which mirrors the increased density registered toward the southern part of the town.\textsuperscript{\cite{NRPDen}} As demonstrated in Figure~\ref{fig:nrstats}{\color{red}b)}, the distribution of household sizes and the mean size is in good agreement with Census data. As evidenced in Figures~\ref{fig:nrstats}{\color{red}c)} and~\ref{fig:nrstats}{\color{red}d)}, we closely match the distribution of employed family members and the overall age distribution, accurately resolving several age groups. As reported in Table~\ref{tab:nrstats}, we are successful in maintaining similar percentages of households with agents who are above 60 years old. This aspect is particularly important for realistic predictions of COVID-19, which is significantly more severe and fatal among the older population. Aside from age, the modeled community preserves the ratio of families with children and of single-parent households, as shown in Table~\ref{tab:nrstats}.

\begin{table}
\centering
\caption{Modeled properties of the population and US Census data for New Rochelle, NY.}
\begin{tabular}{lcc}
\hline
Category & Model & US Census value \\
\hline
Mean household size & 2.77 & 2.71\\
Households with one or \\more individuals 60 years and older & 47.0\% & 42.3\% \\
Families with children & 39.1\% & 34.2\% \\
Single parent families & 18.7\% & 25.0\% \\
\hline
\label{tab:nrstats}
\end{tabular}
\end{table}

All children aged 5-17 were assigned to schools of appropriate levels. The number of students in a school was proportional to school size based on the data from the National Center for Education Statistics.\textsuperscript{\cite{NCES}} Accordingly, the initial distribution of students in the model, representing time before the pandemic, was set so that a portion of children younger than five years old was placed in daycares, and similarly, 35\% of individuals between 18--21 years old attended the town's higher education institutions. 

A portion of agents older than 16 were allowed to work and study according to a set of rules based on the estimated workplace and school sizes (see Section S3 of the Supporting Information). Agents' workplace distribution was generated using the US Census data on the percentage of the population working in a given industry while the number of employees at hospitals, schools, and retirement homes was estimated. The initial numbers of retirement homes residents were estimated based on institution size. The initial number of hospital patients with a condition other than COVID-19 was set to occupy one-sixth of the total available beds. The exact steps for creating the population from building information and census data are outlined in Section S3 of the Supporting Information.

\subsection{COVID-19 data sources}
The number of total and active New Rochelle cases were collected manually from official reports and videos available for Westchester County.\textsuperscript{\cite{WestchesterCountyPR}} The mortality and testing statistics were obtained from the Official Twitter account of the Westchester County,\textsuperscript{\cite{WestchesterCountyPR}} and the  New York State Department of Health,\textsuperscript{\cite{NYSTesting}} respectively, for Westchester County as a whole, and were then scaled to the population of New Rochelle. All the data is available through our repository (\url{https://github.com/Dynamical-Systems-Laboratory/NR-ABM-population}).

\section{COVID-19 ABM with testing and treatment}
\subsection{Model overview}
During a day, agents transition between different locations (identified in the New Rochelle database: households, workplaces, schools, retirement homes, and hospitals). In each of these locations, they can interact with other agents, thereby supporting the transmission of COVID-19. Agents can be healthy, undergoing testing, or be under treatment. We assume that the town is isolated, such that the agents cannot leave the town, and new agents cannot enter during the simulation.

\begin{figure}
  \centering
  \includegraphics[width=0.7\textwidth]{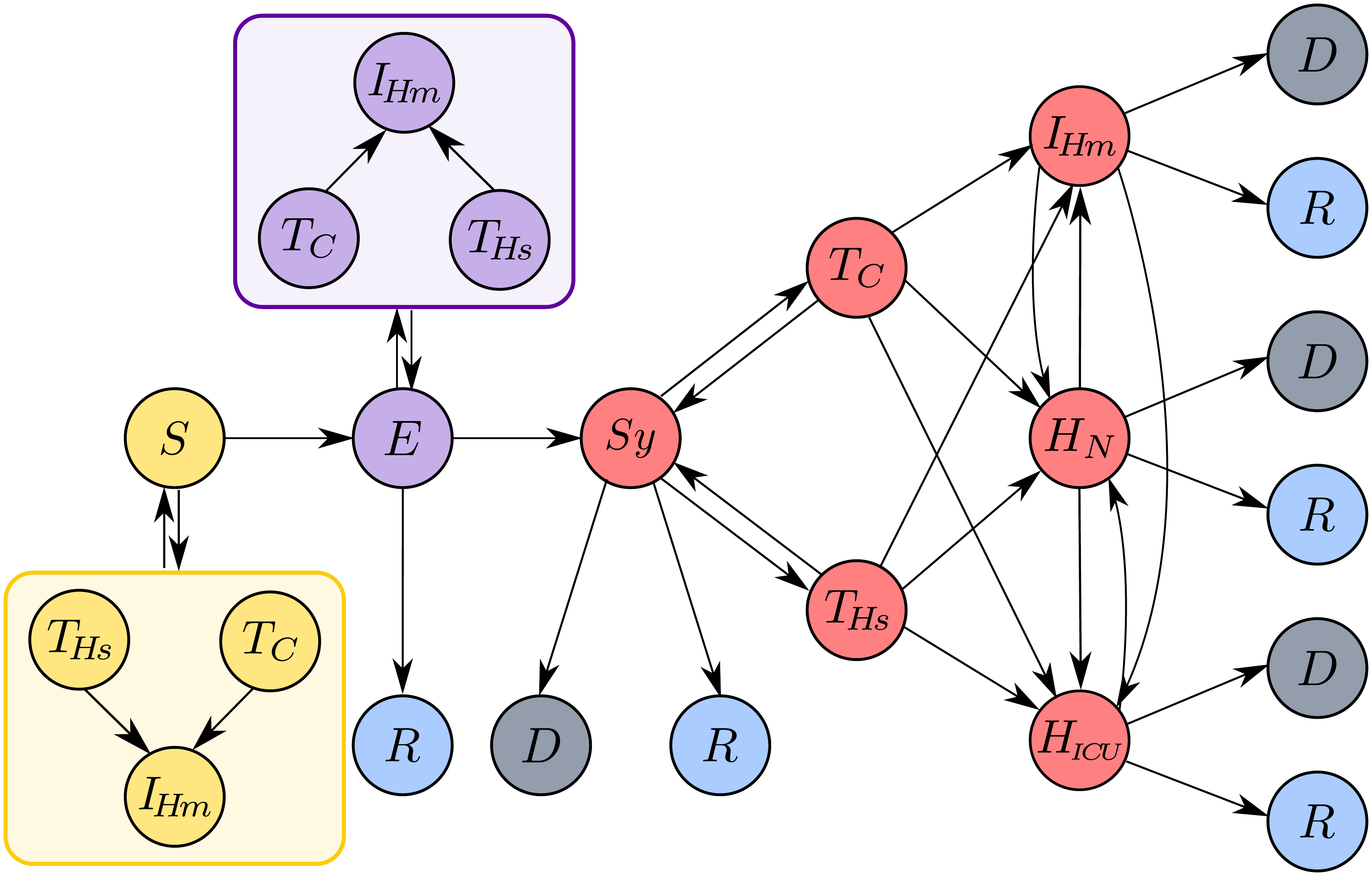}
  \caption{Schematic representation of modeled agent states and their possible transitions. Agent in the model can be in one of the following states: susceptible ($S$); exposed ($E$); symptomatic ($Sy$); removed - dead ($D$); removed - healthy/recovered ($R$); Agents in different states can undergo testing in a test car ($T_C$), or a hospital ($T_{Hs}$) after which they can be treated through home isolation ($I_{Hm}$), normal hospitalization ($H_N$), or hospitalization in an intensive care unit, ICU ($H_{ICU}$). In addition to symptomatic agents, exposed agents and agents who have COVID-19-like symptoms but are not infected can also be tested. Except for the symptomatic agents, all positive test results, including false positives, will lead to home isolation.}
  \label{fig:abmtes}
\end{figure}

Motivated by~\cite{ferguson2005strategies,ferguson2006strategies,ferguson2020impact}, the COVID-19 progression model consists of five states: susceptible ($S$), exposed ---which includes infectious individuals who have not yet developed symptoms--- ($E$), infectious-symptomatic (${Sy}$), removed-healed (${R}$), and removed-dead (${D}$); with a detailed outline of states, their variants, and transitions as shown in Figure~\ref{fig:abmtes}. Upon infection, susceptible agents become exposed (${E}$) and remain so for a latency period. Following the COVID-19 infectiousness profile, we assume that exposed agents are not infectious during the initial part of the latency period. When the latency period is over, exposed agents develop symptoms and become infectious-symptomatic (${Sy}$).\textsuperscript{\cite{aleta2020modeling}} Some exposed agents may recover without ever developing symptoms (that is, they are asymptomatic); in this case, their latency period is extended to match the expected COVID-19 recovery period.

Exposed agents and agents showing symptoms ---whether from COVID-19 or another condition--- have the possibility of being tested which can be performed either in a hospital ($T_{Hs}$) or a car ($T_C$). Testing in a hospital carries the possibility of infecting hospital staff and patients with a condition other than COVID-19. Additionally, if the agent is COVID-19-negative, then the agent is at risk of becoming infected during testing. We assume that drive-through testing does not carry the risk of infection based on the work of Upham.\textsuperscript{\cite{DriveTesting}} The outcomes of a test can be true positive or negative or false positive or negative. 

After testing, agents are assigned treatment from one of the following three treatment types: home isolation ($I_{Hm}$), routine hospitalization ($H_N$), and hospitalization in an ICU ($H_{ICU}$). An exposed agent who has tested positive for COVID-19 will always undergo home isolation until developing symptoms, at which point their treatment can potentially change to $H_N$ or $H_{ICU}$. In contrast, a symptomatic agent can be assigned to one of the three treatments. A symptomatic agent can transition between different treatment types during the course of the disease. All infected agents are removed through either recovery ($R$) or death ($D$). A removed agent no longer contributes to the spread of the infection. 

A symptomatic agent who is untested will not undergo treatment, and their removal is determined following similar rules to the tested agents. However, an untested agent requiring ICU treatment has an increased probability of dying, to account for the higher mortality in the absence of a diagnosis. While the agents undergoing testing are routinely quarantined, those who are not deemed to be tested retain their normal activities. However, when developing symptoms, these agents will refrain from going to work or school, thereby reducing the contribution to contagion in public areas. 

Susceptible agents can have symptoms similar to COVID-19 due to non-COVID-19 diseases, such as a common cold or seasonal influenza.\textsuperscript{\cite{CDCFluCOVID}} We assume that this population is constant throughout the duration of the simulation. Since these susceptible agents are suspected of having COVID-19, they can undergo testing and thus introduce two additional elements in our model: i) they increase the count of people being tested, and therefore the burden on testing sites, and ii) they can contract COVID-19 upon interacting with infected people at the testing site. Finally, such agents can be erroneously tested positive during their recovery from the non-COVID-19 disease. However, after recovery, these agents are still susceptible to COVID-19.

To simulate realistic COVID-19 epidemic conditions, we introduce school closures, a state-wide lockdown, and the three reopening phases I, II, and III. School closure is modeled by omitting the contribution of schools to the transmission of COVID-19. Similarly, lockdown and reopening phases are characterized by tuning the contributions of the workplaces to the transmission. The number of tests performed per day can be time-dependent, following real practices.\textsuperscript{\cite{NYSTesting}}

\subsection{COVID-19 transmission dynamics}

The proposed epidemiological model consists of COVID-19 transmission through agents interacting at their residences and public places. The agents reside in households, retirement homes, or hospitals when being treated for conditions other than COVID-19. They can also attend schools and go to work; employees of schools, retirement homes, and hospitals are modeled explicitly for their high-risk and critical role. Following the original work of Ferguson, there is no distinction between times of the day, for example day versus night. At each simulation step, $\Delta t$, an agent may contract the disease or infect other agents at home, school, workplace, or hospital.\textsuperscript{\cite{ferguson2005strategies, ferguson2006strategies}} For example, if a susceptible agent is a high school student who also works part-time, their probability of being infected with COVID-19 is computed based on their contacts at their school, workplace, and household. When agents are being tested or hospitalized, they do not infect those in their households, schools, or workplaces.

The model comprises a set of agents $\mathcal N=\{1,...,n\}$ and a set of locations $\mathcal L=\{1,...,L\}$. According to the town database, an agent is associated with a subset of locations determined by the model input. Formally, we define a set of functions $f_q:\mathcal N\to\mathcal L$, with  $q\in\{\house,\work,\school,\retire,\hsp\}$, so that function $f_q$ associates each agent $i\in\mathcal V$ to the corresponding location $\ell\in\mathcal L$ of type $q$. The types of locations are households ($\house$), workplaces ($\work$), schools ($\school$), retirement homes ($\retire$), and hospitals ($\hsp$). Note that each agent may not be associated to all the types of locations. To denote that agent $i$ is not associated to a location of type $q$, we write $f_q(i)=\emptyset$. We denote by $n_\ell$ the number of agents associated with location $\ell$.

At every simulation step (of duration $\Delta t$), infected agents assigned to location $\ell$ contribute to the probability of infection for susceptible agents at that location. Specifically, the probability that an agent $i$ that is susceptible at time $t$ becomes infected in the following time-step is equal to
\begin{equation}
    p_i(t) := 1 - e^{-\Delta t \Lambda_i\left(t\right)},
\label{eq:pii}
\end{equation}
where the non-negative time-varying parameter $\Lambda_i\left(t\right)$ quantifies the contagion risk at all the locations associated with the agent and it is equal to
\begin{equation}
\begin{aligned}
    \Lambda_i\left(t\right) := \lambda_{\house,f_\house(i)}\left(t\right) + \lambda_{\work,f_\work(i)}\left(t\right) + \lambda_{\school,f_\school{i}}\left(t\right) + \lambda_{\retire,f_{\retire}(i)}\left(t\right) + \lambda_{\hsp,f_{\hsp}(i)}\left(t\right),
\end{aligned}
\label{eq:lam}
\end{equation}
where each contribution represents the so-called infectiousness function of each type of location associated with agent $i$, with the understanding that $\lambda_{q,\emptyset}=0$, that is, if agent $i$ is not associated with any location of type $q$, than locations of type $q$ do not contribute to the agent's contagion risk.

The infectiousness function of a location $\ell$ of type $q$ at time $t$, $\lambda_{q,\ell}\left(t\right)$, due to all the agents (indexed by $k$) at that location, is  defined as 
\begin{equation}
\begin{aligned}
    \lambda_{q,\ell}\left(t\right) := \frac{1}{{n_\ell^{\alpha_q}}}\sum_{k=1}^{n_\ell}
    \left(E_k\rho_k\beta_{q,k} 
    + Sy_k\psi_\ell c_k\rho_k\beta_{q,k}\right).
\end{aligned}
\label{eq:str}
\end{equation}
The sum is performed over the $n_\ell$ agents who are associated with location $\ell$ and it represents a weighted ratio between the number of exposed and infected agents and all the agents at the location; $E_k$ is an indicator function that is equal to $1$ if agent $k$ is in the exposed ($E$) and has become infectious and $0$ otherwise; $Sy_k$ is equal to $1$ if agent $k$ is symptomatic ($Sy$) and $0$ otherwise. The parameter $\rho_k\geq 0$ models variability in infectiousness among the agents; $c_k > 1$ is a factor that measures the increased infectiousness of a symptomatic agent compared to an exposed one; $\alpha_q\leq 1$ is a size scaling parameter (less than one for households and one otherwise); $\psi_\ell\in[0,1]$ is an absenteeism correction for workplaces and schools, which is used to model reduction of agent presence upon developing symptoms; $\beta_{q,k}\geq 0$ is a transmission rate that generally depends on the type of location $q$ and on the activity of agent $k$ at that particular location (for example, the transmission rate for an agent who is being tested at a hospital is different from an agent who works there). Further details of the model are provided in Section S1 of the Supporting Information. 

When an agent becomes exposed, they undergo an incubation period. The latency of the incubation is drawn from a log-normal distribution, which allows for the possibility of some agents to not spread the virus until they develop symptoms. Once the incubation ends, an agent transitions from exposed to symptomatic. The model allows for a portion of exposed agents to recover without symptoms (commonly referred to as asymptomatic individuals) by including their recovery time within their incubation period.

\subsection{Testing}\label{subsec:covtes}
Both exposed and symptomatic agents can undergo testing for COVID-19 according to two different probabilities. When an agent is scheduled to be tested, they are placed under home isolation and randomly assigned to a testing location --- a drive-through or a  hospital. We assume that the test is performed for a fixed amount of time after the decision to be tested, and similarly that the result of a test appears after a fixed delay following the test. The test result can be either positive (true or false) or negative (true or false), with negative results causing the agent to return from home isolation to the community. An exposed agent confirmed positive for COVID-19 remains in home isolation, while a symptomatic one is given an initial treatment.

Testing is performed differently for exposed hospital employees and patients originally admitted for non-COVID-19-related diagnosis (such as a car accident or cancer treatment). These agents do not undergo home isolation, and their testing is always performed in the hospitals they work or reside in, without the option of a test car. The symptomatic hospital staff is home isolated prior to receiving the test results, while the non-COVID-19 patients stay in the hospital and, upon developing COVID-19 symptoms, they are counted among hospitalized COVID-19 cases. This fine level of detail is needed to capture evidence of extensive COVID-19 spreading in the early stage of the pandemic in hospitals.\textsuperscript{\cite{wang2020reasons,gan2020preventing,schwartz2020protecting}} After confirming COVID-19, agents are assigned treatment. 

With the exception of hospital employees and patients who develop disease symptoms, the model does not apply any explicit contact tracing. Instead, case detection is implemented in an average sense. Whether an agent will be tested is determined by stochastic sampling of a uniform distribution, followed by a comparison with testing prevalence at that time.

\subsection{Treatment}\label{subsec:covtre}

When a symptomatic agent is confirmed COVID-19 positive, they are assigned to one of the three formal treatments according to a probabilistic mechanism: home isolation, regular hospitalization, and hospitalization in an ICU.\textsuperscript{\cite{ferguson2020impact,hinch2020openabm}} Afterward, the agent can change treatment types depending on their recovery status and clinically observed COVID-19 progression. The agent's initial treatment is chosen based on the probability of normal hospitalization and hospitalization in an ICU obtained from clinical data depending on their age.\textsuperscript{\cite{ferguson2020impact,verity2020estimates}} In both cases, agents are assigned to a random hospital. If hospitalized in an ICU, their recovery status is recomputed based on an agent's probability of dying in an ICU.\textsuperscript{\cite{ferguson2020impact}} All other agents are placed in home isolation. 

Treatment changes include all possible transfers except direct transfer from an ICU to home isolation, as outlined in Figure~\ref{fig:abmtes}. In our model, whether an agent dies or recovers is determined upfront when the agent develops symptoms of the disease. Treatment transitions are closely related to an agent's future recovery outlooks. An agent originally determined to die and treated in an ICU will die in the ICU.\textsuperscript{\cite{richardson2020presenting}} Upon recovering, agents will be transferred to normal hospitalization after an amount of time decided a priori.\textsuperscript{\cite{ferguson2020impact}} Any dying agent who was previously confirmed COVID-19 positive would be placed in an ICU for a predetermined number of days before death.\textsuperscript{\cite{salje2020estimating}} A recovering agent initially hospitalized outside an ICU can become home isolated if their recovery time exceeds their hospital stay.\textsuperscript{\cite{richardson2020presenting}} Finally, an agent recovering while isolated at home can become hospitalized for a certain amount of time, a commonly observed course for the disease.\textsuperscript{\cite{richardson2020presenting,linton2020epidemiological}} 

\subsection{Initialization and vaccination}\label{subsec:covini}

At the beginning of the simulation, we initialize the entire population as susceptible, assuming no prior immunity in a virtually COVID-19-free population. Then, a predefined number of agents are assigned the exposed health state. These agents can only be tested after developing symptoms. 

Part of the susceptible population can also be vaccinated and thus becomes immune. The vaccines are distributed by two modes: i) randomly throughout the entire population, or ii) to a specific type of agent, such as healthcare workers or retirement home employees.

\subsection{Agent removal}\label{subsec:covrem}

Disease progression can have two possible outcomes: recovery or death. The outcome is determined using age-based mortality data, the agent's treatment requirements, and current testing prevalence. An agent's mortality also depends on whether they are tested and receive proper medical attention.\textsuperscript{\cite{salje2020estimating,xie2020critical,phua2020intensive,pulla2020counts,oxley2020large}} Specifically, while asymptomatic agents always recover, we distinguish between two events that can occur to symptomatic agents and influence their probability of dying. 

The model decides if an agent needs ICU care upon exhibiting symptoms. An agent who needs an ICU will be admitted upon being tested. Not all the agents who need an ICU will be admitted to one; those who are not tested, and therefore not diagnosed, will die. Among the agents who do not need an ICU, a fraction may still die, for example, due to
heart failure, stroke, or a rapid decline in condition; some of these individuals will die in their homes, but some others will formally be admitted to ICU, despite not needing it, based on the agent's expected lifetime. 

All these probabilities are available from empirical observations except the probability of dying without the need of an ICU. In order to estimate this quantity, we perform the following calculations. We expand the overall probability that a symptomatic agent dies, $P(D|Sy)$, using the law of total probability with respect to the conditioning on whether the symptomatic agent needs ICU treatment, $N$, and the event that the symptomatic agent does not need an ICU, $\overline{N}$, 
\begin{equation}\label{eq:PD}
    P(D|Sy)
        =P(D|N)P(N)
        +P(D| \overline{N})P(\overline{N})\,.
\end{equation}
 By rearranging Equation~\eqref{eq:PD}, we obtain the following closed-form expression for the required probability
\begin{equation}
        P( D|\overline{N}) = \frac{P(D|Sy) - P(D| N) P(N)}{1 - P(N)}\,.
\label{eq:dnicu}
\end{equation} 
In the following, we derive the three expressions for $P(D|Sy)$, $P(D|N)$, and $P(N)$, which are needed to compute the formula.

First, the probability of dying for symptomatic agents, $P(D|Sy)$, is inferred from the infection fatality ratio ($IFR$) available in the literature.\textsuperscript{\cite{ferguson2020impact}}  Since the $IFR$ is based on serology-informed estimations, it reflects the probability of dying for an infected agents (regardless of whether the agent is symptomatic or not). Assuming that asymptomatic agents do not die, we can compute the overall time-averaged probability that a symptomatic agent dies by re-scaling the $IFR$ by the probability of developing symptoms once contracting COVID-19,  $P(Sy|\textrm{CoV})$, obtaining
\begin{equation}
        P({D|Sy}) = \frac{I FR}{P(Sy|\textrm{CoV})}\,.
\label{eq:dtot}
\end{equation}

Second, the probability of dying if an agent needs an ICU, $P(D|N)$, in Equation~\eqref{eq:dnicu} is computed depending on whether they are tested once they become symptomatic. Specifically, by means of the law of total probability with respect to the conditioning on the event $T$, we have
\begin{equation}\label{eq:test_no_test}
P\left(D|N\right)=1-\left(1-P(D|N,T)\right)\cdot P\left(T|Sy\right)\,,
\end{equation}
where $P\left(T|Sy\right)$ is the probability that a symptomatic agent is tested. Since the $IFR$ used in Equation~\ref{eq:dtot} is a temporal average over the entire duration of the pandemic,  the probability $P\left(T|Sy\right)$ is also estimated as an average over the entire duration of the pandemic, from data reported in Supplementary Table S9.

Third, the probability that a symptomatic agent needs ICU care, $P(N)$, is computed from empirical data on the probability that an agent needs to be hospitalized, $P(H)$, and the probability that symptomatic hospitalized agents need ICU care, $P(N|H)$, yielding
\begin{equation}\label{eq:PICU}
    P(N) =P(N|H) P(H)\,,
\end{equation}
where these empirical data are reported in Supplementary Table S7.

Finally, the required probability is computed by substituting Equations \eqref{eq:dtot}, \eqref{eq:test_no_test}, and~\eqref{eq:PICU} into Equation \eqref{eq:dnicu}.

When an agent dies, they are removed from all public places, hospitals, or current residences. Recovered agents become active in public locations, and if hospitalized, return to their households or retirement homes. Recovered agents who were previously hospitalized for a condition other than COVID-19 are readmitted to the hospital.

\subsection{Susceptible agents with COVID-19-like symptoms}
Susceptible agents with COVID-19-like symptoms do not exist in the model until the onset of testing. Once testing efforts begin, these agents are assigned as a fraction of those who are still susceptible. A portion of these agents will undergo testing and receive either a false positive or a true negative result. Therefore, the probability of a susceptible agent with COVID-19-like symptoms to be tested,  $P(T, Sy, \overline{\textrm{CoV}})$, is given as
\begin{equation}
    P(T, Sy, \overline{\textrm{CoV}}) =\big( P\left(\textrm{true negative}|Sy\right)+P\left(\textrm{false positive}|Sy \right)\big )P(T).
\end{equation}
We assume that the probability of getting a false negative among COVID-19-infected symptomatic agents when the epidemic prevalence is low is negligible. \textsuperscript{\cite{LAPH,FNeg}} We can then approximate the probability of a true negative by the probability of receiving a negative result.
 
If scheduled to be tested, the agent is assigned a test time and a testing site (either a randomly chosen hospital or a drive-through test). When the test occurs, the time is selected from a Gamma distribution to avoid these agents undergoing home isolation and testing simultaneously. Similar to the procedure for an infected agent undergoing testing, the agent displaying COVID-19-like symptoms is placed under home isolation for a certain amount of time before testing occurs. Home isolation lasts until the agent is confirmed negative or reaching ``recovery" after a false-positive result. The duration of the home isolation before the test and the subsequent wait time for results is the same for these agents as for the infected ones. 

If an agent with COVID-19-like symptoms contracts COVID-19, they become an exposed agent. To maintain a fixed fraction of such agents in the population, a new susceptible agent is then randomly chosen to take their place, provided such agents are still present in the population.

\subsection{Lockdown and reopening events}
Our model provides options to simulate school closure, lockdown, and three reopening phases, 1, 2, and 3. School closures are simulated by zeroing the transmission rates of students and employees. The business closure and reopening are implemented through user-defined reduction or increase of the initial workplace transmission rates, respectively. The model also allows for adjusting the absenteeism correction of a workplace, that is, $\psi_\ell$ in Equation~\eqref{eq:str}, to a lockdown value, valid through the reopening phases. The transmission parameters for households, hospitals, and retirement homes remain unchanged throughout the simulation.
\subsection{Model parameters}

Parameters originate from several sources: established literature data used in other \\
ABMs\textsuperscript{\cite{ferguson2005strategies,ferguson2006strategies,ajelli2010comparing,ferguson2020impact}}, clinical data on COVID-19\textsuperscript{\cite{ferguson2020impact,richardson2020presenting,linton2020epidemiological,lauer2020incubation}}, and information from a clinical consultant who is part of the team. In addition, due to the lack of concrete data, some of our parameters are informed estimates, in line with the current understanding of COVID-19 from scientific literature and the media. Furthermore, some parameter types are identified from reported data through model calibration. These latter parameters are the number of initially infected agents, time-varying testing prevalence, COVID-19 transmission changes following closures and reopening phases, and asymptomatic agents' age distribution. Parameters, data sources, and assumptions are listed and indicated in Section S2 of the Supporting Information.

There are four groups of model parameters: COVID-19 transmission dynamics parameters, testing parameters, parameters related to closure and reopening events, and other parameters, all listed as Tables in the Supporting Information. Transmission dynamics originates from the COVID-19 agent-based model in Ferguson et al.\textsuperscript{\cite{ferguson2020impact}} While not explicitly stated, the transmission dynamics parameters used therein mirror those previously developed for influenza by the same research group.\textsuperscript{} Such a choice is justified since COVID-19 is a respiratory disease that spreads\cite{ferguson2005strategies, ferguson2006strategies} similarly to influenza. However, to make the transmission rates more representative, we further scale them by the ratio of reproductive numbers, $R_0$, for these two diseases. $R_0$ represents the average number of secondary infections directly caused by a single infected individual.\textsuperscript{\cite{brauer2011mathematical}} Following analogous models and procedures, $R_0$ for COVID-19 was estimated to be $2.4$,\textsuperscript{\cite{ferguson2020impact}} while for influenza it was reported as $R_0=1.7$, resulting in a scaling factor of $1.41$.\textsuperscript{\cite{ferguson2006strategies}}

Hospital-related transmission rates are calculated by scaling equivalent non-hospital rates with data from our clinical consultant in Italy. Specifically, we use the fact that there was a 7.2\% increase in infection among hospital employees in a given week in Italy compared to a 3.7\% increase in the general population. Thus, we use a ratio of these percentages as our scaling factor to multiply a base rate of choice. The base rate for a hospital employee is the workplace rate, and for an agent hospitalized as a non-COVID-19 patient, it is the household rate. Other hospital rates are set relative to these following personal communication with the clinical consultant in Italy.

Hospitalization duration in the model is derived from the literature,\textsuperscript{\cite{ferguson2020impact}} and linearly scaled by a factor of $0.39$ according to the data in the paper by Richardson et al.\textsuperscript{\cite{richardson2020presenting}} which are specific to the geographic region considered in this work. Specifically, the study by Richardson et al.\textsuperscript{\cite{richardson2020presenting}} provides actual hospitalization duration in New York City, though without distinguishing between ICU and non-ICU treatment, and relative lengths of these two. Hence, we use the ratio of the total hospital treatment duration reported in Richardson et al.\textsuperscript{\cite{richardson2020presenting}} and Ferguson et al.\textsuperscript{\cite{ferguson2020impact}} to obtain locally realistic hospitalization periods.

\section{Model validation}\label{sec:resmod}
To demonstrate our platform's viability, we simulated the spread of COVID-19 from February 22\textsuperscript{nd} to July 14\textsuperscript{th}, 2020, from the onset of the epidemic to phase three of the reopening. In particular, this window includes the period in which all the schools in the town of New Rochelle were closed (March 13\textsuperscript{th}, 2020), followed by the restrictive state-wide lockdown during which only essential businesses, such as grocery stores, were allowed to operate.

In our calibration, we used officially reported data on the total number of detected infections, number of people currently infected with the disease, and the total number of fatalities. From the cumulative number of cases and mortality, we extracted the number of new cases and deaths reported each week. To calibrate upon this dataset, we varied the initial number of infected agents, the percentage of tested population, the reduction in workplace transmission rates during the lockdown and its subsequent increase during the reopening periods, and age-dependent fractions of asymptomatic agents as is summarized in Section S2 of the Supporting Information. Parameters were manually initialized, while the testing percentages were later refined using simplex optimization in MATLAB via the \texttt{fminsearch} function. Testing prevalence was set to vary with time, mimicking the actual testing practices in the region.\textsuperscript{\cite{NYSTesting}} In other words, the fraction of infected individuals who could be tested was time-dependent. To match it, we used the data on newly confirmed cases every week during the simulation period, computed from total detected cases. We performed 100 realizations of the simulation, randomly selecting a fixed number of initially infected agents each time. All the parameters used in the simulations are listed in Section S2 of the Supporting Information. The computational performance of the model is summarized in Section S4 of the Supporting Information; the implementation is fairly efficient and approachable for general use, with 600 steps (150 days) of the simulation taking less than 30s on a standard laptop.

\begin{figure}
  \centering
  \includegraphics[width=\textwidth]{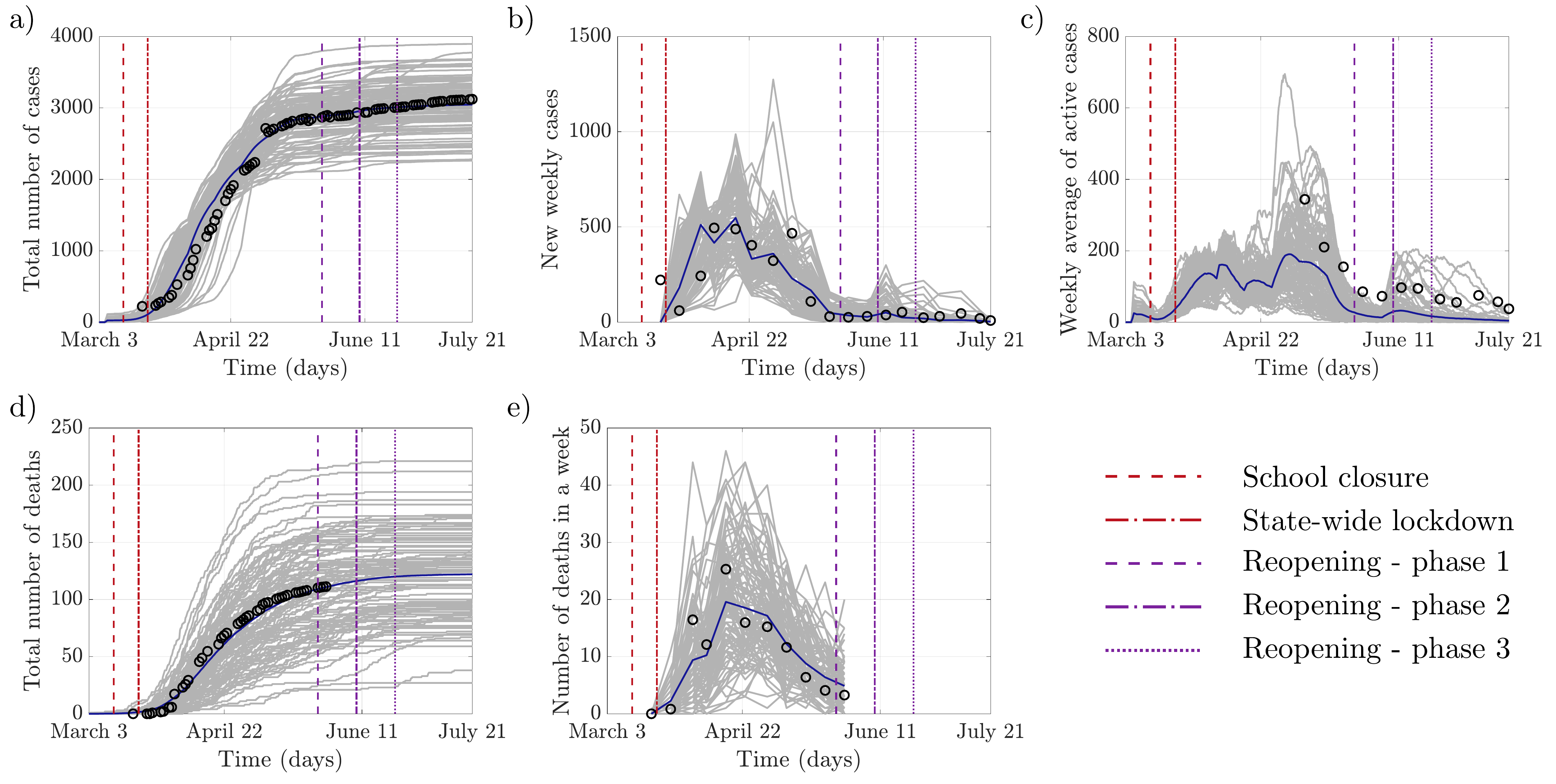}
  \caption{Comparison of the modeled COVID-19 epidemic and officially reported data: a) The cumulative number of infections; b) New infections detected within a week; c) Active cases averaged over each week; d) The total number of deaths;  e) Number of deaths in each week.  The grey lines represent each of the simulation's 100 realizations, the blue line is the average value, and black circles are the reported data.}
  \label{fig:comp}
\end{figure}

\begin{figure}
  \centering
  \includegraphics[width=0.7\textwidth]{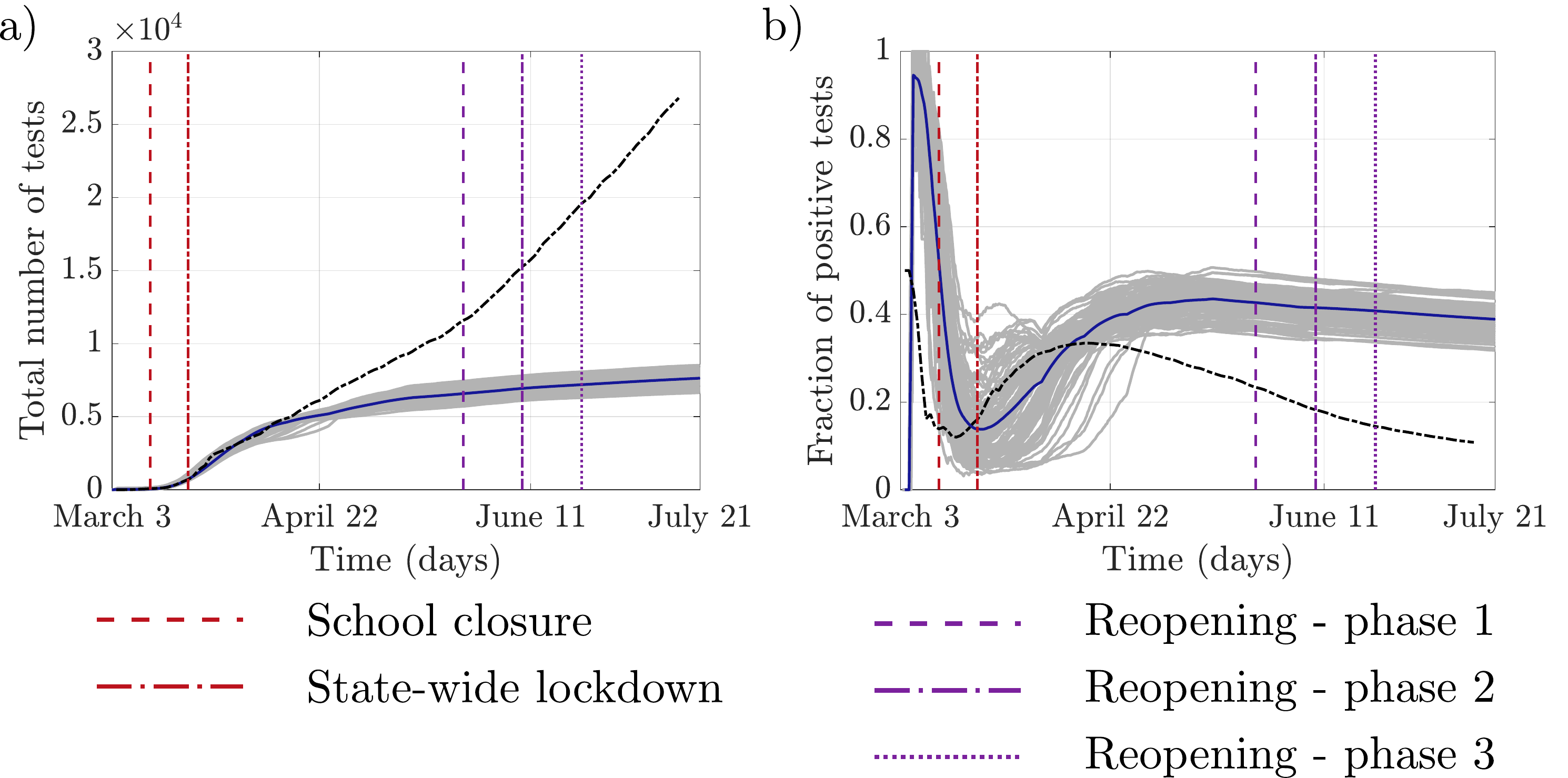}
  \caption{Comparison of modeled and reported testing practices: a) The total number of performed tests; b) Fraction of positive test results, including false positives: the grey lines represent each of 100 realizations of the simulation, the blue line is the average value, and dashed black line is the reported data.}
  \label{fig:tests}
\end{figure}

Figures~\ref{fig:comp} and~\ref{fig:tests} show the results of the validation. Figure~\ref{fig:comp} compares the model output with real data along with five different metrics: i) the total number of cases, ii) the number of new cases, iii) the weekly average of active cases, iv) the total number of deaths, and v) the number of deaths in a week. The total number of cases was calculated as the number of agents who tested positive, including false positives and those who died without treatment. The number of new weekly cases was calculated as the weekly increase in the total number of cases. Working with weekly averages facilitates comparisons by filtering out spurious oscillations from uneven reporting and data collection by the authorities. In our model, the number of active cases was computed as the number of agents undergoing treatment confirmed positive or false positive. These were compared directly to the reported weekly average. The number of deaths includes both treated and undetected, untreated agents,
assuming that COVID-19 would be confirmed in individuals who have died, regardless of their testing status. The number of deaths per week was computed as the weekly rise in total fatalities.  

A comparison of the total number of cases shows good agreement between the model and real data obtained from official outlets of Westchester County.\textsuperscript{\cite{WestchesterCountyPR}} Similarly, the number of new cases is well predicted by our model. Looking closely, however, the model has a smooth progression of the disease compared to a sudden high number of initial cases in the real data. We note that our model imparts a simplistic scenario of the testing practice, whereby an aggressive contact tracing followed the town's initial case detection.\textsuperscript{\cite{NRNews}} This likely resulted in a large difference in the number of initial cases in our model versus the real data. Simulating this particular scenario is currently outside the scope of our model.

In the case of weekly averages of active cases, the model reasonably matches real data trends. At the same time, the mean value predicted by the model is slightly lower than the reported values, likely due to longer recovery times of COVID-19 patients than utilized in the model. We note that compared to the total number of reported cases, which provide information about new infections, the number of active cases also includes the process of recovery. Regarding the number of deaths, the reported values were obtained for the entire Westchester County and scaled down to New Rochelle, proportionally to its population. Here too, we find a close agreement between simulated and real data.

Figure~\ref{fig:tests} compares the total number of tests and positivity in our simulations with the available data from local testing practices. Similarly to the number of deaths, this data was reported at the county level and was scaled down to match the population of New Rochelle for a meaningful comparison. Our results indicate reasonable agreement in the early phases of the epidemic with discrepancies later on. The lower number of total tests in our model is due to the rule used for testing, which is based on the total number of exposed and symptomatic individuals. In contrast, in reality, testing was ramped up to include the general population. The only susceptible agents who can be tested in the model, in its present incarnation,  are those exhibiting COVID-19-like symptoms. The fraction of these individuals is low and chosen upfront in the model. This trend is also visible in the positivity values, whereby we find an inflated positivity in our simulations by a factor of four compared to the real data, again due to limited testing in our model. This difference shows that negative testing outcomes do not affect the general number of cases, as evidenced by the model agreement on the number of cases and deaths.

\begin{figure}
  \centering
  \includegraphics[width=0.7\textwidth]{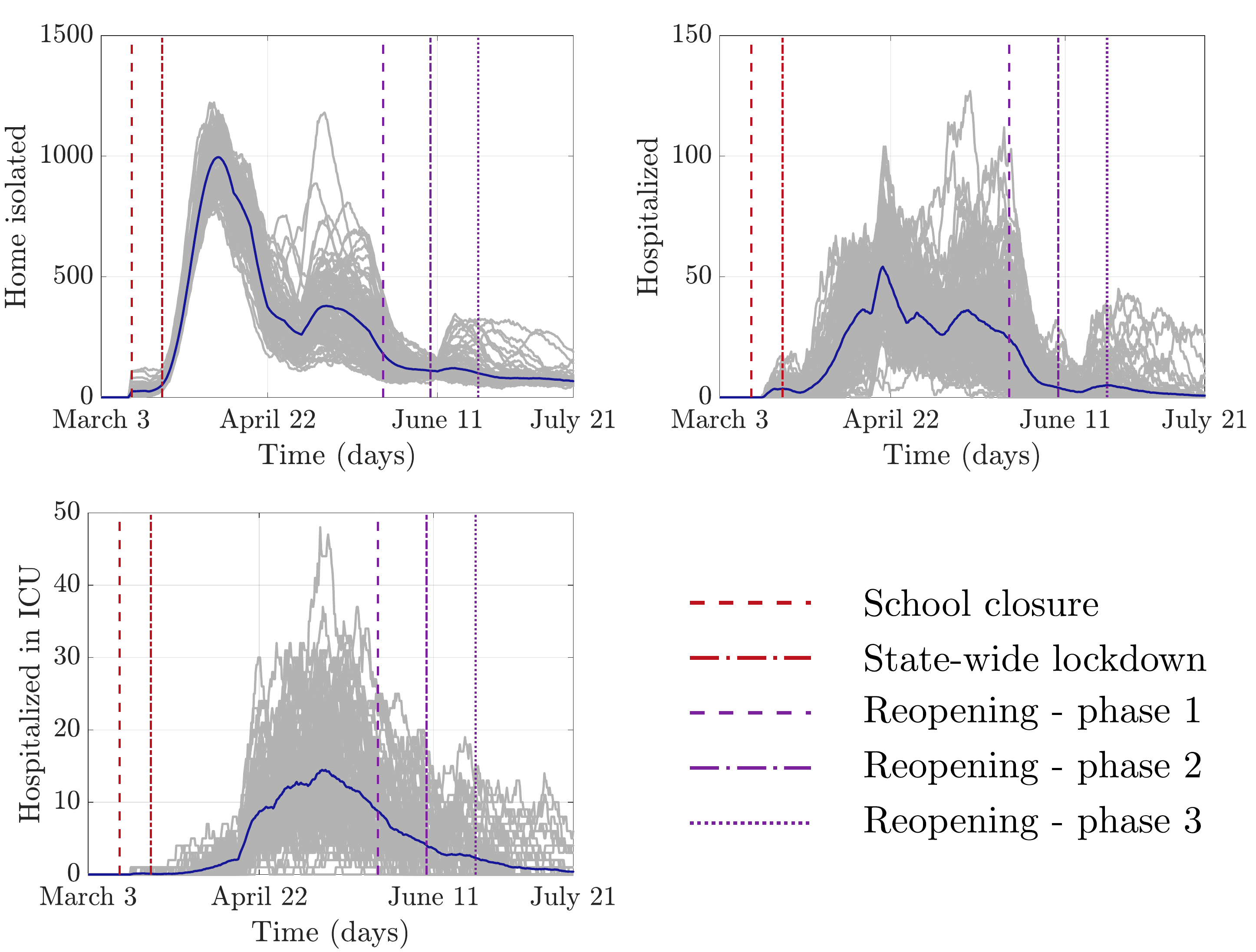}
  \caption{The number of agents undergoing each of the three treatment types at different time points in the simulation. The grey lines represent 100 realizations of the simulation; the blue line is the average value.}
  \label{fig:trt}
\end{figure}

Figure~\ref{fig:trt} shows the number of agents who were undergoing a given treatment type. The number of agents isolated at home comprised individuals who were waiting for a test or test results. The prevalence of each type of treatment qualitatively matches the general distribution of cases. Home isolated individuals constituted the bulk of infected agents, followed by hospitalized individuals, and finally, a few hospitalized in ICUs.

According to the New York State Department of Health, the New Rochelle hospital has 211 general and 12 ICU beds at its disposal under normal circumstances.\textsuperscript{\cite{NYSHealth}} In all realizations of our model, the number of hospitalized agents was always below the reported normal bed volume. In the model, the ICU demand was on average within standard hospital capacities, but in many simulations exceeded it two or even threefold. Given the expansion of hospital capacity as a response to COVID-19\textsuperscript{\cite{NRMtHsp}} and in the absence of reliable data, we consider this agreement reasonable.

\section{Vaccination study} 

To demonstrate the value of our platform, we performed a comparative analysis of different vaccination strategies. Specifically, we evaluated the effect of vaccinating only high-risk groups of individuals, hospital, school, or retirement home employees, or retirement home residents and compare the results to a random immunization across the entire population.

The time period of this prospective study was aligned with the first wave of the epidemic, making the previously calibrated model the basis for the prediction. None of the parameters were changed in this study with respect to the earlier validation. The only differences in this vaccination study were the absence of school closures and any form of lockdown. In this context, the study also investigated the consequences of leaving schools and non-essential businesses open throughout the first wave of the epidemic upon the availability of a vaccine.

Vaccination was implemented in the simulation on March 2\textsuperscript{nd}, simultaneously with the beginning of testing in New Rochelle. All the vaccines were distributed simultaneously, granting full protection against the disease. By then, some of the agents were already infected and were therefore excluded from vaccination. Susceptible agents with COVID-19-like symptoms were not vaccinated either, in an attempt to maintain an approximately fixed number of such agents in the simulation. We performed six sets of simulations with vaccinations of: i) hospital employees only, ii) school employees only, iii) retirement home employees only, iv) retirement home residents only,  v) randomly selected fraction of the population, with the same size as the number of hospital employees, and vi) about a quarter of the town, corresponding to ten times the number of hospital employees.

\begin{figure}
  \centering
  \includegraphics[width=\textwidth]{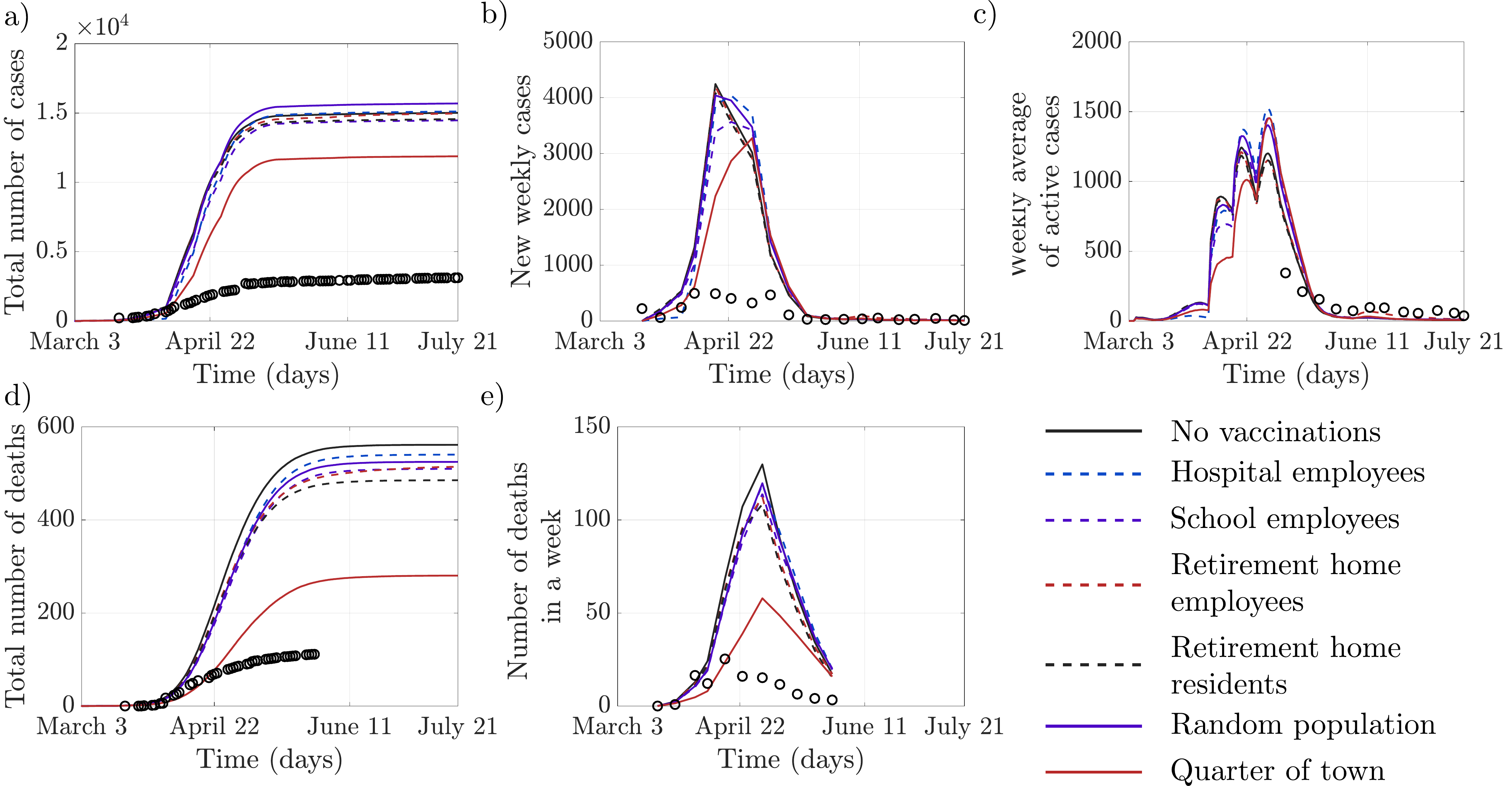}
  \caption{These figures show the spread of COVID-19 epidemics across a range of ``what-if'' scenarios, when no lockdowns or closures were set in place, and vaccinations were administered selectively. Only means are reported for visual clarity; also, officially reported data (black circles) are presented for reference. In the scenario where the only vaccinated agents were ones from specific high-risk groups the vaccination covered on average 2,201 hospital employees ($\pm 11$), 7,577 school employees ($\pm 11$), 494 retirement home employees ($\pm 2$), and 1,397 retirement home residents ($\pm 5$).The number in parentheses indicates the standard deviation. In the fifth scenario, the number of people in the general population  vaccinated was equivalent to the number of vaccinated hospital employees. A tenfold increase in the number of vaccinated individuals among the general population was considered in the third scenario.}
  \label{fig:vac}
\end{figure}

Figure~\ref{fig:vac} shows the predictions from these six ``what-if'' scenarios. The importance of closures is evident, with numbers of infections and fatalities exceeding reality many times. The vaccination of hospital employees resulted in only minor differences compared with the vaccination of an equivalent number of individuals among the general population. Similar observations can be made about targeted immunization of other vulnerable groups of agents. Significant differences only occur in mortality when vaccinating the elderly residents of retirement homes. Although both, targeted and random, approaches had some effect on COVID-19 spread, massive immunization was the only truly impactful strategy. This finding is consistent with ``herd immunity'' predictions where effective containment of COVID-19 can only be achieved with the large majority of the population acquiring immunity.\textsuperscript{\cite{randolph2020herd}}

\section{Conclusions}

Until widespread vaccination efforts are underway, maintaining a balance between safety and normalcy during the current COVID-19 crisis requires the use of non-pharmaceutical prevention measures as well as efficient detection strategies. The large number of testing strategies, unknowns, and high levels of uncertainties of this epidemic calls for the principled use of predictive computational models, potentially informing policy-making with respect to widespread vaccination efforts.

In this work, we proposed a high-resolution ABM of COVID-19, developed for granular simulations of a small city or town, where each individual is explicitly modeled. We introduced several elements of novelty with respect to state of the art on ABM, including i) different testing strategies in hospitals and drive-throughs; ii) time variations in testing
prevalence; iii) multiple types of treatment, from home isolation to hospitalization in an
ICU; iv) the presence of susceptible agents who have COVID-19-like symptoms due to other
infections; v) explicit modeling of employees of hospitals, schools, and retirement homes;
vi) school and business closures and reopenings; vii) comprehensive model calibration with officially reported data; and viii) incorporation of expert knowledge from the field.

We applied our model to the US town of New Rochelle, where one of the first COVID-19 outbreaks in the country took place. Using an in-house, detailed database of building locations, public and residential, and Census data, we created a geographically and statistically accurate representation of the town and its population. We demonstrated the possibility of accurately capturing the first wave of the COVID-19 epidemic in the town.

As New Rochelle is a representative US small town, we believe that our validated model can serve as an analysis platform for numerous similar towns across the entire country, many currently facing the COVID-19 crisis. To illustrate the model's value in analyzing prospective ``what-if''  questions, we performed an immunization study in which we evaluated several vaccination strategies of future importance. In particular, we compared the impact of vaccination of select group of vulnerable individuals, including school employees, retirement home employees and residents, and the totality of the two thousand hospital employees in the town, a randomly selected group of two thousand individuals, and twenty thousand randomly selected individuals out of the eighty thousand people living in New Rochelle. Our results suggest that prioritizing vaccination of high-risk individuals has a marginal effect on the count of COVID-19 deaths. Predictably, a much more significant improvement is registered when a quarter of the town is vaccinated. Importantly, the benefits of the restrictive measures in place during the first wave greatly surpass those from any of these selective vaccination scenarios.    

While undoubtedly useful, our model bears several limitations. First, the model lacks explicit agent mobility and random contacts, which manifest in a faster decline of the epidemic near the end of the simulation. The original model by Ferguson et al.\textsuperscript{\cite{ferguson2005strategies, ferguson2006strategies}}, serving as the basis for our own, had an additional term to the model disease spreading through random contacts in the community. However, these contacts were based on commute and travel data at the level of the entire country, which is not directly applicable to the problem at hand of a small town. Simulating truly random interactions using a contact network approach similar to Hinch et al.\textsuperscript{\cite{hinch2020openabm}} may offer an alternative, which will be part of our future work. Along these lines, the impact of local travel and commute can further be included in the model by integrating traffic flow simulations.\textsuperscript{\cite{frias2011agent,barbosa2018human,zheng2013primer}}

In addition to mobility and random contacts, our model does not include testing the general population, leading to possible under-detection of cases in later reopening stages. While the model allows for testing uninfected individuals with symptoms, massive community testing is needed to align its outcomes with reality in later phases of the epidemic. Testing of general population along with random interactions is also expected to highlight the effects of different testing strategies already encoded within the model. Combined with contact tracing options, testing of general population is part of the next step in our platform's development.

Another limitation of our approach is in the modeling of hospitals in terms of their workforce and capacity. Specifically, we assume that infections of hospital employees do not trigger changes in the treatment of hospital patients and that hospitals have infinite beds and ICUs. Finally, we do not explicitly account for the use of personal protective equipment (PPEs), such as face coverings, and social distancing of agents. While these measures are included indirectly through reduction of disease transmission during the lock-down and reopening phases, the ability of specific agents to protect themselves from the contagion would improve the granularity of the model and add a further realistic element, at the expenses of the computational burden. Advantageous impact of PPEs and distancing can be introduced to both susceptible and infected agents in a similar way to agent's current infectiousness variability, and this will be one of our goals in the nearest future. 

Despite these limitations, our model matched real data very closely as the epidemic progressed through its initial stages. This correspondence allowed us to prospect and analyze alternative scenarios for COVID-19, in which vaccination was accessible right at the onset of the first wave. Beyond the timely study of vaccination strategies, our model can be adapted to explore a range of pressing problems that are ahead of us by interested users who can directly modify our open-source platform. For example, the model can be swiftly adapted to describe the concurrent spread of influenza with COVID-19, which is expected to exacerbate the impact of second and third waves. Likewise, the model can provide clear and quantitative support to the long-debated recommendations regarding the need to avoid large gatherings and always use masks.

\medskip
\textbf{Supporting information} \par 
The database and the code for generating the town's population are available at \url{https://github.com/Dynamical-Systems-Laboratory/NR-ABM-population}. The ABM code is available at \url{https://github.com/Dynamical-Systems-Laboratory/ABM-COVID-DSL}. 

\medskip
\textbf{Author contributions}\\
Conceptualization - AT, SB, ZPJ, AR, MP; methodology - AT, LZ, SB, ZPJ, AR, MP; software - AT; validation - AT; formal analysis - AT, BB, LZ; investigation - all the authors; resources - MP; data curation - JH; writing – original draft preparation - AT, BB, LZ; writing – review and editing - JH, SB, EC, ZPJ, AR, MP; visualization - AT; supervision - SB, EC, ZPJ, AR, MP; project administration - MP; funding acquisition -  SB, ZPJ, AR, MP. 

\medskip
\textbf{Acknowledgements} \par 
We would like to acknowledge Kyle C Payen for collecting and post-processing the officially reported COVID-19 data, Malav Thakore for help with writing and verifying the manuscript draft, and Anna Sawulska for help with  designing the graphical abstract. This work was partially supported by National Science Foundation (CMMI-1561134,  CMMI-2027990, and CMMI-2027988), Compagnia di San Paolo, MAECI (``Mac2Mic"), the European Research Council (ERC-CoG-771687), and the Netherlands Organisation for Scientific Research (NWO-vidi-14134). 

\medskip
\textbf{Conflict of Interest} \par 
The authors declare no conﬂict of interest.

\pagebreak

\cleardoublepage



\section*{Supplementary material}

# S1 Detailed formulation of COVID-19 transmission dynamics

In this section, we detail the analytical expressions used for computation of COVID-19 infectiousness functions in our model, as outlined in Section 3.2 in the main article.

The probability that an agent $i$ who is susceptible at discrete time $t$ will be infected at the following time-step of the simulation depends on the contagion risk $\Lambda_i(t)$, which consists of the sum of the infectiousness functions of all the locations that the agent is associated with. Following the notation used in the main article, we denote by $f_q(i)$ the location of type $q$ associated with agent $i$, where $f_q(i) = \emptyset$ if no location of type $q$ is associated with agent $i$. The expression for the infection risk is represented by the following equation (that is, Equation (2) in the main article):

$$\Lambda_i(t) := \lambda_{\text{H},f_\text{H}(i)}(t) + \lambda_{\text{W},f_\text{W}(i)}(t) + \lambda_{\text{S},f_\text{S}i}(t) + \lambda_{\text{Rh},f_{\text{Rh}}(i)}(t) + \lambda_{\text{Hsp},f_{\text{Hsp}}(i)}(t), \tag{S1}$$

where $\lambda_{\text{H},f_\text{H}(i)}(t)$ is the risk associated with the household in which agent $i$ lives, $\lambda_{\text{W},f_\text{W}(i)}(t)$ is the risk associated with agent $i$'s workplace, $\lambda_{\text{S},f_\text{S}(i)}(t)$ is the risk associated with agent $i$'s school, $\lambda_{\text{Rh},f_{\text{Rh}}(i)}(t)$ is the risk associated with agent $i$'s retirement home, and $\lambda_{\text{Hsp},f_{\text{Hsp}}(i)}(t)$ is the risk associated with agent $i$'s hospital. According to the agent type and their health state, only some of the terms in Equation (S1) may be different from zero, since an agent may not be associated with a location of type $q$. Table S1 lists, for each type and health state of agents, the locations that are associated to the agent and give a nonzero contribution to their corresponding Equation (S1).

The infectiousness function of a location $\ell$ of type $q$ at time $t$ is defined by Equation (3) of the main article, that is,

$$\lambda_{q,\ell}(t) := \frac{1}{n_\ell^{\alpha_q}} \sum_{k=1}^{n_\ell} \left(E_k \rho_k \beta_{q,k} + Sy_k \psi_\ell c_k \rho_k \beta_{q,k}\right). \tag{S2}$$

Therein, specific elements may be kept or omitted, depending on location characteristics. Equation (S2) can be further detailed for different types of locations. Equations (S3)–(S7) refer to household, workplaces, schools, retirement homes, and hospitals, respectively. In these equations, to discriminate among different contributions to the formation of the infectiousness functions by agents belonging to different categories, we use indicator functions denoted as $E_k$ and $Sy_k$, that indicate a generic category and the function takes value one if the agent $k$ belongs to that category, and zero otherwise. All the indicator functions used are listed in Table S2.

In all the Equations (S3)–(S7), $\rho_k \geq 0$ denotes the variability in infectiousness among the agents; $c_k > 1$ is a factor that measures the increased infectiousness of a symptomatic agent compared to an exposed one; and $n_\ell$ is the number of agents at that location.

**Household.** The infectiousness function of a household $\ell$ consists of two separate contributions of untreated agents that live in the household and agents that are home isolated in the household, denoted by subscripts Ut and Ih, respectively, yielding the following expression:

$$\lambda_{\text{H},\ell} = \frac{1}{n_\ell^\alpha} \sum_{k=1}^{n_\ell} \left(E_{k,\text{Ut}} \rho_k \beta_{\text{H,Ut}} + E_{k,\text{Ih}} \rho_k \beta_{\text{H,Ih}} + Sy_{k,\text{Ut}} c_k \rho_k \beta_{\text{H,Ut}} + Sy_{k,\text{Ih}} c_k \rho_k \beta_{\text{H,Ih}}\right), \tag{S3}$$

where $\alpha$ is the size scaling parameter, $\beta_{h,\text{Ut}}$ is the transmission rate of untreated agents, and $\beta_{h,\text{Ih}}$ is the transmission rate of agents who are undergoing home isolation, as part of the testing procedure or subsequent treatment.

**Workplaces.** For a generic workplace $\ell$, that is, one that is not a school, a retirement home, or a hospital, the infectiousness function is defined as

$$\lambda_{\text{W},\ell} = \frac{1}{n_\ell} \sum_{k=1}^{n_\ell} \left(E_{k,\text{W}} \rho_k \beta_\text{W} + Sy_{k,\text{W}} c_k \rho_k \psi_\text{W} \beta_\text{W}\right), \tag{S4}$$

where $\beta_\text{W}$ is the transmission rate of the workplace and $\psi_\text{W}$ is the absenteeism correction. The latter denotes the portion of agents who will still be present at their workplace, regardless of having COVID-19 symptoms. For example, $\psi_\text{W} = 0.20$ means that 20% of symptomatic agents will still go to work. If an individual $i$ is

| Type of susceptible agent | $\lambda_{\text{H}}$ | $\lambda_{\text{W}}$ | $\lambda_{\text{S1}}$ | $\lambda_{\text{S2}}$ | $\lambda_{\text{Rh}}$ | $\lambda_{\text{Hsp}}$ |
|---|---|---|---|---|---|---|
| Hospital employee | ● | | | | | ● |
| Hospital employee attending school $s1$ | ● | | ● | | | ● |
| Non-COVID-19 patient in a hospital | | | | | | ● |
| Agent with COVID-19-like symptoms at the time of testing | | | | | | ● |
| Home isolated retirement home resident with COVID-19-like symptoms | | | | | ● | |
| Home isolated agent with COVID-19-like symptoms | ● | | | | | |
| Agent with COVID-19-like symptoms who works | ● | ● | | | | |
| Agent with COVID-19-like symptoms who works at a school $s1$ | ● | | ● | | | |
| Agent with COVID-19-like symptoms who works at a school $s1$ and attends school $s2$ ($s2$ can be $s1$) | ● | | ● | ● | | |
| Agent with COVID-19-like symptoms who works at a retirement home and attends school $s1$ | ● | | ● | | ● | |
| Resident of a retirement home | | | | | ● | |
| Employee of a retirement home | ● | | | | ● | |
| Employee of a retirement home attending school $s1$ | ● | | ● | | ● | |
| School employee at school $s1$ | ● | | ● | | | |
| School employee at school $s1$ attending school $s2$ ($s2$ can be $s1$) | ● | | ● | ● | | |
| Agent who works at a general workplace | ● | ● | | | | |
| Agent who works at a general workplace and attends school $s1$ | ● | ● | ● | | | |

Table S1: Contributions to the formation of the infectiousness functions for each agent type and health state. The dots indicate the terms that are not zero for each agent category. Subscripts to the functions $\lambda$ correspond to: H - household, W - workplace, S - school, S1, S2 — potentially different schools for school employees if they are both working at a school and attending one, Rh - retirement home, and Hsp - hospital. All agents reside in either a household or a retirement home, unless they are hospital patients treated for a condition different than COVID-19 (e.g., undergoing a surgery). Agents who have COVID-19-like symptoms but are not infected are in home isolation when waiting for a scheduled test, the test results, and if tested false positive.

| Indicator | Description |
| --- | --- |
| $E_{k,\text{Ut}}$ | Exposed agent in a household who is untreated |
| $E_{k,\text{Ih}}$ | Exposed agent in a household who is home isolated |
| $Sy_{k,\text{Ut}}$ | Symptomatic agent in a household who is untreated |
| $Sy_{k,\text{Ih}}$ | Symptomatic agent in a household who is home isolated |
| $E_{k,\text{W}}$ | Exposed agent at a workplace |
| $Sy_{k,\text{W}}$ | Symptomatic agent at a workplace |
| $E_{k,\text{Emp}}$ | Exposed agent who is a school employee |
| $E_{k,\text{St}}$ | Exposed student |
| $Sy_{k,\text{Emp}}$ | Symptomatic agent who is a school employee |
| $Sy_{k,\text{St}}$ | Symptomatic student |
| $E_{k,\text{RhEmp}}$ | Exposed retirement home employee |
| $E_{k,\text{RhUt}}$ | Exposed retirement home resident who is untreated |
| $E_{k,\text{RhIh}}$ | Exposed retirement home resident who is home isolated |
| $Sy_{k,\text{RhEmp}}$ | Symptomatic retirement home employee |
| $Sy_{k,\text{RhUt}}$ | Symptomatic retirement home resident who is not treated |
| $Sy_{k,\text{RhIh}}$ | Symptomatic retirement home resident who is home isolated |
| $E_{k,\text{HspEmp}}$ | Exposed hospital employee |
| $E_{k,\text{Pt}}$ | Exposed patient previously admitted with a condition different than COVID-19 |
| $E_{k,\text{Tst}}$ | Exposed agent getting tested at that time-step |
| $Sy_{k,\text{Pt}}$ | Symptomatic patient previously admitted with a condition different than COVID-19 |
| $Sy_{k,\text{Hn}}$ | Symptomatic agent routinely hospitalized with COVID-19 |
| $Sy_{k,\text{ICU}}$ | Symptomatic agent hospitalized in an ICU with COVID-19 |
| $Sy_{k,\text{Tst}}$ | Symptomatic agent getting tested at that time-step |

Table S2: Summary of indicator functions used in Equation (S3)–Equation (S7). The value of each indicator is 1 if the agent falls into the category in the description, and 0 otherwise. An indicator of 1 means that this agent will be included in the summation, contributing to the infectiousness function of the location.

associated with a workplace that is either a school, a retirement home, or a hospital, then the term $\lambda_{\text{W},i}(t)$ in the expression of $\lambda_i(t)$ in Equation S1 is computed according to the corresponding formula for a school, a retirement home, or a hospital in Equations (S5)–(S7), detailed in the following.

**Schools.** The infectiousness function of a school $\ell$ accounts for both presences of the employees, denoted with the subscript Emp, and students, denoted with the subscript St, according to:

$$\lambda_{\text{S},\ell} = \frac{1}{n_\ell} \sum_{k=1}^{n_\ell} (E_{k,\text{Emp}} \rho_k \beta_{\text{S},\text{Emp}} + E_{k,\text{St}} \rho_k \beta_{\text{S},\text{St}} + Sy_{k,\text{Emp}} c_k \rho_k \psi_{\text{Emp}} \beta_{\text{S},\text{Emp}} \\ + Sy_{k,\text{St}} c_k \rho_k \psi_{\text{St},\ell} \beta_{\text{S},\text{St}}). \tag{S5}$$

An employee contributes to the school infectiousness function with a transmission rate of $\beta_{\text{S},\text{Emp}}$, while a student with a transmission rate $\beta_{\text{S},\text{St}}$. The two categories of agents may, in principle, have different absenteeism corrections, $\psi_{\text{Emp}}$ and $\psi_{\text{St},\ell}$, respectively. Note that the absenteeism correction for students, $\psi_{\text{St},\ell}$, depends on the location $\ell$, since it can vary across different school types: daycare, primary school, middle school, high school, and college.

**Retirement homes.** In a retirement home, $\ell$, the infection may spread through employees, denoted with the subscript Emp, and residents, denoted with the subscript Rh. Similarly to a household, residents can be either untreated, Ut, or home isolated, Ih. The infectiousness function has the following expression:

$$\lambda_{\text{Rh},\ell} = \frac{1}{n_\ell} \sum_{k=1}^{n_\ell} (E_{k,\text{RhEmp}} \rho_k \beta_{\text{Rh},\text{Emp}} + E_{k,\text{RhUt}} \rho_k \beta_{\text{Rh},\text{RhUt}} + E_{k,\text{RhIh}} \rho_k \beta_{\text{Rh},\text{RhIh}} \\ + Sy_{k,\text{RhEmp}} c_k \rho_k \psi_{\text{Rh},\text{Emp}} \beta_{\text{Rh},\text{Emp}} + Sy_{k,\text{RhUt}} c_k \rho_k \beta_{\text{Rh},\text{RhUt}} \\ + Sy_{k,\text{RhIh}} c_k \rho_k \beta_{\text{Rh},\text{RhIh}}), \tag{S6}$$

where employees' transmission rate is $\beta_{\text{Rh,Emp}}$ while the transmission rates of an untreated and a home-isolated resident are $\beta_{\text{Rh,RhUt}}$ and $\beta_{\text{Rh,RhIh}}$, respectively; $\psi_{\text{Rh,Emp}}$ is the absenteeism correction for employees.

**Hospitals.** Computation of the infectiousness function of a hospital $\ell$ should account for several types of agents: i) hospital employees (denoted by the subscript Emp), ii) patients who were admitted with a condition other than COVID-19 and got infected while hospitalized (subscript Pt), iii) patients who are routinely hospitalized with COVID-19 (subscript Hn), iv) patients hospitalized with COVID-19 in an intensive care unit, ICU (subscript ICU), and v) infected agents who are getting tested for COVID-19 in the hospital (subscript Tst). Contributions of all these agents yield the following expression:

$$\lambda_{\text{Hsp},\ell} = \frac{1}{n_\ell} \sum_{k=1}^{n_\ell} (E_{k,\text{HspEmp}}\rho_k \beta_{\text{Hsp,Emp}} + E_{k,\text{Pt}}\rho_k \beta_{\text{Hsp,Pt}} + E_{k,\text{Tst}}\rho_k \beta_{\text{Hsp,Tst}} \\ + Sy_{k,\text{Pt}}c_k\rho_k \beta_{\text{Hsp,Pt}} + Sy_{k,\text{Hn}}c_k\rho_k \beta_{\text{Hsp,Ih}} \\ + Sy_{k,\text{ICU}}c_k\rho_k \beta_{\text{Hsp,ICU}} + Sy_{k,\text{Tst}}c_k\rho_k \beta_{\text{Hsp,Tst}}). \tag{S7}$$

Equation (S7) is derived upon a number of model assumptions. First, hospital employees do not come to work once they develop disease symptoms. This is modeled through the lack of symptomatic contribution and any absenteeism correction for hospital employees. Thus, a hospital employee will contribute to $\lambda_{\text{Hsp},\ell}$ only as an exposed agent, with a transmission rate of $\beta_{\text{Hsp,Emp}}$. Such complete absenteeism holds even before the population develops awareness of the disease, reflecting specific hospital conditions as a workplace. However, before the COVID-19 detection period starts, a symptomatic hospital employee may still contribute to the infectiousness function in a school if they are attending one. A patient admitted with a condition other than COVID-19 can contribute both as exposed and as symptomatic agents, with a transmission rate of $\beta_{\text{Hsp,Pt}}$. After disease awareness and testing begin, hospital employees and patients are always tested upon developing symptoms. While all agents who are being tested, including the symptomatic hospital employees, are placed in home isolation during the testing procedure, hospital patients remain in the hospital, but their status and transmission rate change to an agent hospitalized with COVID-19. Furthermore, agents who are routinely hospitalized with a COVID-19 diagnosis can only be symptomatic and are characterized by a transmission rate of $\beta_{\text{Hsp,Ih}}$. The same holds for agents hospitalized in an ICU, having a transmission rate of $\beta_{\text{Hsp,ICU}}$. Finally, any agent tested in the hospital contributes to the contagion at the test's time step with a transmission rate of $\beta_{\text{Hsp,Tst}}$.

## S2 Model parameters

Here, we list all the parameters on which our model relies and reproduce the results in Sections 4 and 5 of the main article.

Parameters are briefly discussed in Section 3.9 of the main article and are here presented grouped into four categories: COVID-19 transmission dynamics parameters (Table S3), testing parameters (Table S4), parameters related to closures and reopenings (Table S5), and other parameters (Table S6). Additionally, Table S7 outlines the treatment and mortality statistics, Table S8 shows the age-dependent portion of agents who will not develop symptoms during the course of the disease, and Table S9 records the fractions of exposed and symptomatic agents tested as a function of time. References are included, and the assumptions made are clearly stated. The distributions of symptoms onset to death and hospitalization from [1] were digitized using the free software WebPlotDigitizer [2].

| Parameter | Value | References |
|---|---|---|
| Severity correction, $c_k$ | 2.0 | [3, 4] |
| Infectiousness variability, $\rho_k$ | Gamma distribution with mean 1, shape parameter 0.25, scale parameter 4 | [5] |
| Household transmission rate - untreated, $\beta_{H,Ut}$, time$^{-1}$ | 0.69 day$^{-1}$ | [6] scaled by 1.41 |
| Household transmission rate - home isolated, $\beta_{H,Ih}$, time$^{-1}$ | 0.48 day$^{-1}$ | Assumption |
| Household size scaling parameter, $\alpha_H$ | 0.8 | [4] |
| Workplace transmission rate, $\beta_W$, time$^{-1}$ | 0.66 day$^{-1}$ | [6] scaled by 1.41 |
| Workplace absenteeism correction, $\psi_w$ | 0.1 | Assumption |
| Retirement home employee transmission rate, $\beta_{Rh,Emp}$, time$^{-1}$ | 0.66 day$^{-1}$ | Assumption |
| Retirement home employee absenteeism correction, $\psi_{Rh,Emp}$ | 0 | Assumption |
| Retirement home resident transmission rate - untreated, $\beta_{Rh,RhUt}$, time$^{-1}$ | 0.69 day$^{-1}$ | Assumption |
| Retirement home resident transmission rate - home isolated, $\beta_{Rh,RhIh}$ | 0.48 day$^{-1}$ | Assumption |
| School student transmission rate, $\beta_{S,St}$, time$^{-1}$ | 1.33 day$^{-1}$ | [6] scaled by 1.41 |
| School employee transmission rate, $\beta_{S,Emp}$, time$^{-1}$ | 0.66 day$^{-1}$ | Assumption |
| School employee absenteeism correction, $\psi_{Emp,S}$ | 0.1 | Assumption |
| Student absenteeism correction, $\psi_{S,St}$ | 0.1 (daycare), 0.1 (primary and middle), 0.1 (high school), 0.1 (college) | Assumption |
| Hospital employee transmission rate, $\beta_{Hsp,Emp}$, time$^{-1}$ | 1.28 day$^{-1}$ | Estimated based on data from a clinical consultant |
| Transmission rate of hospital patients with a condition different than COVID-19, $\beta_{Hsp,Pt}$, time$^{-1}$ | 1.38 day$^{-1}$ | Estimated based on data from a clinical consultant |
| Transmission rate of hospitalized agents, $\beta_{Hsp,Ih}$, time$^{-1}$ | 1.02 day$^{-1}$ | Estimated based on data from a clinical consultant |
| Transmission rate of ICU hospitalized agents, $\beta_{Hsp,ICU}$, time$^{-1}$ | 1.34 day$^{-1}$ | Estimated based on data from a clinical consultant |
| Transmission rate of infected hospital visitors being tested for COVID-19, $\beta_{Hsp,Tst}$, time$^{-1}$ | 1.75 day$^{-1}$ | Estimated based on data from a clinical consultant |

Table S3: COVID-19 transmission parameters. Assumed values were based on discussion with Clinical consultant.

| Parameter | Value | References |
|---|---|---|
| Probability of an agent getting a negative test result, $P\,(\text{negative})$ | 0.89 | [7] |
| Fraction of susceptible agents with COVID-19-like symptoms | 0.0483 | [8] |
| Probability agent is tested in a hospital, $P\,(T_{\text{Hsp}})$ | 0.60 | Data from a clinical consultant |
| Probability of a false negative test, $P\,(\text{false negative})$ | 0.05 | Data from a clinical consultant |
| Probability of a false positive test, $P\,(\text{false positive})$ | 0.05 | Data from a clinical consultant |
| Wait time for the test, time | 2.25 days | Mid point of data from a clinical consultant |
| Time from test to results, time | 1.75 days | Mid point of data from a clinical consultant |
| Time the testing procedure starts for a susceptible agents with COVID-19-like symptoms | $-10 + \Gamma\,(x)$, with $\Gamma\,(x)$ a value from Gamma distribution with a shape parameter 25.4621, and scale parameter 1.4301 | Assumption |

Table S4: Testing parameters.

| Parameter | Value | References |
|---|---|---|
| Start of the simulation | February 22$^{\text{nd}}$, 2020 (day 0) | [9] |
| Start of data collection | March 2$^{\text{nd}}$, 2020 (day 9) | A day before first confirmed case [10] |
| Start of testing | March 2$^{\text{nd}}$, 2020 (day 9) | A day before first confirmed case [10] |
| School closure | March 13$^{\text{th}}$, 2020 (day 20) | [11] |
| State-wide lockdown | March 22$^{\text{nd}}$, 2020 (day 29) | [12] |
| Reopening - phase I | May 26$^{\text{th}}$, 2020 (day 94) | [13] |
| Reopening - phase II | June 9$^{\text{th}}$, 2020 (day 108) | [14] |
| Reopening - phase III | June 23$^{\text{rd}}$, 2020 (day 122) | [15] |
| Workplace transmission rate reduction - lockdown, $\beta_w$ | 0.1 | Calibrated parameter |
| Workplace transmission rate reduction - reopening phase I, $\beta_W$ | 0.35 | Calibrated parameter |
| Workplace transmission rate reduction - reopening phase II, $\beta_W$ | 0.4 | Calibrated parameter |
| Workplace transmission rate reduction - reopening phase III, $\beta_W$ | 0.5 | Calibrated parameter |
| Workplace lockdown and reopening absenteeism correction, $\psi_W$ | 0 | Assumption |

Table S5: Parameters related to closures and reopening.

| Parameter | Value | References |
|---|---|---|
| Latency period, $\tau_E$, time | log-normal distribution with 1.621 mean and 0.418 standard deviation, days | [16] |
| Number of days between infection and infectiousness, time | 4.6 days | [5] |
| Probability of death in an ICU, $P(D|N,T)$ | 0.5 | [5] |
| Time average of probability a symptomatic agent will be tested, $P(T|Sy)$ | 0.636369 | Average of a calibrated parameter |
| Time spent in a hospital if not admitted to an ICU | 3 days | [5] and [17] |
| Time before death spent in an ICU | 2 days | Assumption |
| Time spent in an ICU if recovering | 4 days | [5] and [17] |
| Time spent in a hospital after an ICU treatment if recovering | 2 days | [5] and [17] |
| Time from the onset of symptoms to recovery | 4.9 days | [5] |
| Time from the onset of symptoms to hospitalization | Gamma distribution with shape parameter 0.7696 and scale parameter 3.4192 | [1] |
| Time from the onset of symptoms to death | Log-normal distribution with 2.6696 mean and 0.4760 standard deviation | [1] |
| Number of initially infected agents | 22 | Calibrated parameter |
| Timestep, $\Delta t$ | 0.25 day | [5] |

Table S6: Other parameters.

| Age group (years) | % symptomatic cases requiring hospitalization | % hospitalized cases requiring ICU | Infection Fatality Ratio |
|---|---|---|---|
| 0-9 | 0.1 | 5.0 | 0.002 |
| 10-19 | 0.3 | 5.0 | 0.006 |
| 20-29 | 1.2 | 5.0 | 0.03 |
| 30-39 | 3.2 | 5.0 | 0.08 |
| 40-49 | 4.9 | 6.3 | 0.15 |
| 50-59 | 10.2 | 12.2 | 0.60 |
| 60-69 | 16.6 | 27.4 | 2.2 |
| 70-79 | 24.3 | 43.2 | 5.1 |
| 80+ | 27.3 | 70.9 | 9.3 |

Table S7: Hospitalization statistics and mortality rates within different age groups as proposed by [5].

| Age group (years) | Fraction |
|---|---|
| 0-12 | 0.9 |
| 13-16 | 0.7 |
| 17+ | 0.5 |

Table S8: Fraction of exposed agents who never develop symptoms (asymptomatic). Values are calibrated consistent with data available in [5].

| Simulation step | Fraction of exposed agents tested | Fraction of symptomatic agents tested |
|---|---|---|
| 9 | 0.8 | 1 |
| 10 | $4.92 \times 10^{-5}$ | $2.48 \times 10^{-5}$ |
| 20 | 0 | $4.85 \times 10^{-5}$ |
| 30 | $4.35 \times 10^{-5}$ | $8.92 \times 10^{-6}$ |
| 40 | $1.46 \times 10^{-5}$ | 0.51 |
| 50 | 0.09 | 0.81 |
| 60 | 0.6 | 0.96 |
| 70 | 0.91 | 1 |
| 80 | 0.098 | 0.76 |
| 90 | $2 \times 10^{-6}$ | 0.65 |
| 100 | 0.8 | 0.95 |

Table S9: Calibrated time-dependent testing prevalence. Simulation steps are counted from February 22$^{\text{nd}}$, 2020 as step 0.

## S3 Sociodemographic model

This section describes the steps for generating the population of New Rochelle, NY, from the in-house town database (https://docs.google.com/spreadsheets/d/1AJAMPvVfxbp5JBC80e0qkgCwemNF2wvC/edit#gid=731369863) and the US Census data. Unless otherwise specified, the Census data is from the 2018 5-year estimates collected in March and April 2020. A brief overview of the steps and data sources is reported in Section 2 of the main article. Unless otherwise specified, all the described operations use the default rounding and floating-point numbers conversion rules of the Python version with which the program is executed.

### S3.1 Generation of households

Households were assigned to the agents using the buildings collected in the database and Census data on the total number of households (https://data.census.gov/cedsci/table?g=1600000US3650617&hidePreview=false&tid=ACSCP5Y2018.CP04&vintage=2017&cid=CP05_2013_001E&t=Housing%3AHousing%20Units). The database distinguishes between single-family houses, townhouses, and multilevel/multi-unit buildings. We assume that one family house represents one household. The information on townhouses includes an estimate of the number of units a townhouse has, and we treat each unit as a single household. For multilevel buildings, we estimate the number of households based on the number of floors registered in the database.

To estimate the number of households in multilevel buildings, we use the following algorithm:

1. **Identify all known housing units**
   Known housing units are single-family homes and townhouses with a clearly identified number of subunits.

2. **Estimate the number of units in apartment complexes**
   The number of units in a given complex, $n_{u,\text{Bld}}$, can be approximated as

$$n_{u,\text{Bld}} = n_{\text{Fl,Bld}} n_{u,\text{Fl}}. \tag{S8}$$

   Here, $n_{\text{Fl,Bld}}$ is the number of floors in the building and $n_{u,\text{Fl}}$ is the approximate number of units per floor calculated as

$$n_{u,\text{Fl}} = \left\lfloor \frac{(N_{u,\text{total}} - N_{u,\text{known}})}{N_{\text{Fl,total}}} \right\rfloor, \tag{S9}$$

   where $N_{u,\text{total}}$ is the total number of units reported in the Census data, $N_{u,\text{known}}$ is the number of known units identified in Step 1, $N_{\text{Fl,total}}$ is the sum of all existing residential floors in New Rochelle, and $\lfloor \cdot \rfloor$ denotes rounding down to the lower integer.

10

3. **Redistribute all the remaining housing units into the multilevel buildings**
   This is done in a round-robin way — each building gets one unit in a circular fashion until no more units are left to distribute.

## S3.2 Generation of public places

Our model distinguishes four types of public places: workplaces, schools, hospitals, and retirement homes. The database on those locations includes geographic coordinates, type of place, short description, Census-defined workplace category under which this location can be classified, and the number of employees. We also identify the number of patients in each hospital, the number of residents in each retirement home, and the number of students in each school, along with the school type.

Number of employees in workplaces other than hospitals are manually adjusted based on Census employment data (https://data.census.gov/cedsci/table?g=1600000US3650617&tid=ACSDP5Y2018.DP03&vintage=2018&layer=VT_2018_160_00_PY_D1&cid=S2501_C01_001E&hidePreview=false&t=Class%20of%20Worker%3AEmployment%3AEmployment%20and%20Labor%20Force%20Status%3AIndustry%3AOccupation) in different workplace categories. The number of employees in all categories, including hospitals, matches the Census's total employed population.

The number of full-time teachers and students at a school is estimated by National Center for Education Statistics (https://nces.ed.gov/globallocator/index.asp?search=1&State=&city=&zipcode=&miles=&itemname=dav&sortby=name&School=1&PrivSchool=1&College=1&Library=1&CS=FB1D5F98). A standalone preschool is considered a daycare, while for schools that combine more types (like primary through high school), both the earliest and latest types are recorded for further processing. Specifically, if there are $N_s$ students in a school and the school combines $N_t$ educational levels (e.g., primary and middle school), each level will be nominally assigned $N_s/N_t$ number of students, rounded down, towards the smaller integer.

Hospital data includes the number of employees and the number of in-patients who, on average, reside in the hospitals. This data is estimated from the records of the New York State Department of Health [18] and the American Hospital Directory (https://www.ahd.com/free_profile/330184/Montefiore_New_Rochelle_Hospital/New_Rochelle/New_York/).

The number of residents in retirement homes is estimated based on the building's size, supported by the websites of specific homes where possible.

## S3.3 Assignment of agents into households

After generating all the public and residential places, we use Census data, and the knowledge derived from it, to create a population of agents and assign it to residential locations as follows:

1. **Create agents in each age group**
   Number of agents in each Census-specified age group is calculated using Census data on population structure (https://data.census.gov/cedsci/table?g=1600000US3650617&tid=ACSST5Y2018.S0101&vintage=2018&layer=VT_2018_160_00_PY_D1&cid=S2501_C01_001E&hidePreview=false&t=Populations%20and%20People) and the total number of agents. The operations round down to the lower integer and then use a round-robin distribution scheme to match the total population reported. The maximum age is assumed to be 100 years old. When an agent is assigned a residence, the count of agents in their age group is decreased.

2. **Assign nursing/retirement home residents and hospital patients**
   Agents aged 75 and older are randomly distributed into retirement homes based on the estimated number of agents in each home. In the same way, agents of any age are randomly assigned as hospital patients.

3. **Exclude vacant households**
   The Census data defines the percent of vacant households (https://data.census.gov/cedsci/table?g=1600000US3650617&tid=ACSCP5Y2018.CP04&vintage=2018&layer=VT_2018_160_00_PY_D1&cid=S2501_C01_001E&hidePreview=false&t=Housing%3AHousing%20Units). The corresponding number of households is excluded and do not have any agents assigned. This number is obtained by rounding down to the lower integer.

4. **Assign the age of a householder**
   Each of the non-vacant households has a householder (head of the household) assigned to it. The householder has to be at least 18 years old, and the distribution follows the Census data on householder age (https://data.census.gov/cedsci/table?g=1600000US3650617&tid=ACSDT5Y2018.B09021&vintage=2010&layer=VT_2018_160_00_PY_D1&cid=S2501_C01_001E&hidePreview=false&t=Housing%3AHousing%20Units). The data is first converted to percentages, after which the number of householders in each age group is calculated. The householder is randomly assigned a specific age in the group and a household.

5. **Select household size**
   Households are assigned sizes based on Census data on household size distribution (https://data.census.gov/cedsci/table?g=1600000US3650617&tid=ACSST5Y2018.S2501&vintage=2018&layer=VT_2018_160_00_PY_D1&cid=S1101_C01_001E&hidePreview=false&t=Housing), following rounding down to the lower integer, finished with a round-robin distribution to match the number of non-vacant households. Each householder is assigned a randomly chosen household size.

6. **Assign remaining household members**
   Households with more than one member are classified as follows:
   - Two parents and children
   - Single parent and children
   - Family with no children (married couple)
   - Person living alone
   - Group of people

   These categories are assigned using probabilities based on percentages from Census data on household structure (https://data.census.gov/cedsci/table?g=1600000US3650617&tid=ACSST5Y2018.S1101&vintage=2018&layer=VT_2018_160_00_PY_D1&cid=S0901_C01_001E&hidePreview=false). Ages of members follow the knowledge base in [3]: there cannot be more than 15 years difference between spouses, and a parent has to be at least 18, and at most 43, years older than the child. Therefore, parents in the model are at most 60 years old, and children are considered as such while below 18 years old. During the entire process, ages are randomly chosen within given bounds, and if the agents are not available anymore in that age interval, an agent with a random, still available age is assigned to that household. Agents remaining after the end of the distribution are randomly assigned to households with four or more members.

   For a household size of 2, it is first determined if the household is a family. If it is a family, they are designated as a married couple or a single parent with a child. Ages are assigned accordingly to the outlined rules. Households with 3 members are assigned either to two parents and a child, one parent and two children, or a random group of adults. Households with four or more members, if not a random group of adults, have one parent and three children or two parents and two children. These households are later randomly assigned to any remaining agents.

   Agents who are 60 years or older are assigned to households or retirement homes, following the Census-reported percentage of households with one or more members of that age range.

### S3.4 Assignment of agents to workplaces and schools

After being distributed in households, agents older than 18 that are not hospital patients or in retirement homes are distributed into workplaces. This operation is performed by assigning agents to workplaces in a sequential fashion, one workplace after another. The maximum working age is set to 75 years. Hospital employees are distributed in the same way except the maximum working age is lower and measures 65 years.

Lastly, agents younger than 22 that are not hospital patients are distributed into schools according to their age,

- < 5 - daycare

- $5 - 10$ - kindergarten and elementary
- $11 - 13$ - middle school
- $14 - 17$ - high school
- $18 - 21$ - college

For each agent, the school that corresponds to their age is chosen randomly from the list of collected schools. Agents are assigned to a school until it reaches its capacity. If a chosen school is already full, the agent is randomly placed in any other school suitable for their age. If there are no more schools to select, agents in daycare and college age are omitted, while others are added randomly to schools in their age group, exceeding nominal capacities.

Since colleges are mostly attended by students from outside New Rochelle who are not counted in the Census, we approximate college attendance instead of introducing outside agents into the model. Specifically, we select a fraction of the population that is $18 - 21$ years old, assumed to pursue higher education, and then evenly distribute it across all colleges in town.

# S4 ABM code details and efficiency

The ABM code is publicly available through our GitHub repository, ABM-COVID-DSL (https://github.com/Dynamical-Systems-Laboratory/ABM-COVID-DSL). The code structure and compilation are explained through an accompanying README.md file. The documentation is created using Doxygen (https://www.doxygen.nl/index.html) — details how to retrieve it are also in the README. There is also a manual and usage examples. The code was extensively tested, with tests and scripts to run them automatically stored in the repository.

We used the exact simulation setup in Section 3 of the main article to estimate the time needed to run our code. The average time out of 100 realizations, each ran for 600 computational steps, was 27.84 s with a standard deviation of 1.95 s. This result shows the applicability of the code for purposes requiring numerous simulations, such as comprehensive parametric studies. The computational time can be further reduced with less data collection.

The time complexity of the code was measured with a synthetic population similar to the one used in the publication. The number of agents was varied from 100 to 1,000,000. Simulations were run for 100 computational steps with no data collection. The average time needed to complete the simulations for each number of agents is indicated in Table S10. For the studied interval, the time complexity was very broadly linear.

Timing and complexity analyses were performed on a MacBook Pro laptop computer with a 2.9 GHz Intel Core i5 processor and 16 GB RAM. The compiler was Apple LLVM version 8.0.0 (clang-800.0.42.1) with C++11 standard and $-O3$ level compiler optimizations.

| Number of agents | Simulation time, s |
|---|---|
| 100 | 0.08 (0.03) |
| 1000 | 0.11 (0.01) |
| 10,000 | 0.57 (0.04) |
| 100,000 | 6.09 (0.45) |
| 1,000,000 | 162.55 (22.65) |

Table S10: Simulation time for different number of agents. Each data point represents an average out of 100 realizations of the code. Standard deviation is indicated in parentheses.

# Supplementary References



\end{document}